\documentclass[preprint,12pt]{elsarticle}

\usepackage[utf8]{inputenc}
\usepackage[T1]{fontenc}
\usepackage{mathtools}
\usepackage{bm}                  
\usepackage{standalone}
\usepackage{multicol}        
\usepackage{multirow}
\usepackage{siunitx}
\usepackage{subcaption}
\usepackage{shellesc}
\usepackage{wrapfig}
\usepackage{amssymb}
\usepackage{amsthm}
\usepackage{amsmath}
\usepackage{xspace}
\usepackage[super]{nth}
\usepackage{scalerel,stackengine}
\usepackage{natbib}
\usepackage{siunitx}
\usepackage{hyperref}

\stackMath
\newcommand\reallywidehat[1]{\savestack{\tmpbox}{\stretchto{\scaleto{\scalerel*[\widthof{\ensuremath{#1}}]{\kern.1pt\mathchar"0362\kern.1pt}{\rule{0ex}{\textheight}}}{\textheight}}{2.4ex}}\stackon[-6.9pt]{#1}{\tmpbox}}
\newcommand{\hwm}{h_{wm}}

\newcommand{\fref}[1]{Fig.~\ref{#1}}
\newcommand{\tref}[1]{Tab.~\ref{#1}}
\newcommand{\sref}[1]{Sec.~\ref{#1}}
\newcommand{\eref}[1]{Eqn.~(\ref{#1})}
\newcommand*{\eg}{e.g.\@\xspace}
\newcommand*{\ie}{i.e.\@\xspace}
\newcommand*{\cf}{cf.\@\xspace}

\newcommand{\tf}[1]{\widehat{#1}}

\journal{arXiv}

\begin{document}

\begin{frontmatter}

\title{Grid-Adaptation for Wall-Modeled Large Eddy Simulation Using Unstructured High-Order Methods}

\author[inst1]{Marcel Blind\corref{cor1}}
\cortext[cor1]{corresponding author}
\ead{blind@iag.uni-stuttgart.de}

\author[inst2]{Ali Berk Kahraman}

\author[inst2]{Johan Larsson}	

\author[inst1,inst3]{Andrea Beck}

\affiliation[inst1]{organization={Institute for Aerodynamics and Gas Dynamics, University of Stuttgart},addressline={Pfaffenwaldring 21},%
city={Stuttgart},%
postcode={70569},%
country={Germany}}

\affiliation[inst2]{organization={Department of Mechanical Engineering, University of Maryland},city={College Park},%
postcode={20742},% 
state={MD},%
country={USA}}

\affiliation[inst3]{organization={Institute of Fluid Dynamics and Thermodynamics, Otto-von-Guericke-University Magdeburg},%
addressline={Universit\"atsplatz 2},% 
city={Magdeburg},%
postcode={39106},% 
country={Germany}}

\begin{abstract}
The accuracy and computational cost of a large eddy simulation are highly dependent on the computational grid.	Building optimal grids manually from \emph{a priori} knowledge is not feasible in most practical use cases; instead, solution-adaptive strategies can provide a robust and cost-efficient method to generate a grid with the desired accuracy. We adapt the grid-adaptation algorithm developed by Toosi and Larsson \cite{toosi:20} to a Discontinuous Galerkin Spectral Elements Method (DGSEM) and show its potential on fully unstructured grids.	The core of the method is the computation of the estimated modeling residual using the polynomial basis functions used in DGSEM, and the averaging of the estimated residual over each element. The final method is assessed in multiple channel flow test cases and for the transonic flow over an airfoil, in both cases making use of mortar interfaces between elements with hanging nodes.	The method is found to be robust and reliable, and to provide solutions at up to 50\% lower cost at comparable accuracy compared to when using human-generated grids.
\end{abstract}


\end{frontmatter}

\section{Introduction}\label{introduction}

As the large eddy simulation (LES) technique gathers increased attention from the engineering community, the open research questions for the method are also changing. While much of the focus in the past was on the development of subgrid scale stress models and numerical methods with limited dissipation, today other issues in the LES method demand more attention. First and foremost among those is the question of how to design a good grid, especially for more complicated geometries. The grid-generation process is still mainly an art that relies on the experience of the user, who needs to incorporate knowledge of numerics and the resolution requirements of flow physics into his/her judgment. This can even lead to a point where different users can create different grids for the same geometry and get different results. Creating an objectively good mesh is crucial for an efficient and resource saving simulation, as well as increasing the dependability of the solutions.
The ultimate objective of this work is to provide a workflow to generate a problem-tailored mesh that is compatible with unstructured high-order methods starting from a very coarse initial mesh generated without any problem-specific insight.

There have been several attempts at devising ways to quantify how well resolved an LES is. Early arguments were based on the ratio of subgrid to viscous dissipation or viscosity (\eg \cite{geurts:02:lesaccuracy,celik:05:lesquality}), but these concepts are meaningful only in the buffer layer of wall-bounded turbulence since LES should be applicable to free shear flows at any Reynolds number. Others have suggested that the sufficiency of a grid in LES should be measured by the ratio of modeled to resolved (or total) turbulence kinetic energy (\cf \cite{jimenez:00, pope:04}), but this has been found to correlate poorly to the known behavior of length scales in wall-bounded flows \cite{toosi:20}. Methods that approximate a local turbulent spectrum have some basis for isotropic flows, but fail for more relevant cases \cite{Flad2016}. The most well-grounded approach to date is that by Toosi and Larsson \cite{toosi:20} which can be viewed as an estimate of the LES modeling residual, \ie, the source term in an error transport equation \cite{toosi:21}.

The objective of the present work is to extend the residual estimate of \cite{toosi:20,toosi:21} to high-order codes and to test the ability of the resulting grid-adaptation method to produce efficient (high accuracy at low cost) grids in a high-order type solver.
In this paper we use the Discontinuous Galerkin (DG) method. Implementation in a DG-type code differs from that in finite-difference or finite-volume codes (where the residual estimate has been tested before) in several ways, including how one performs the low-pass filtering, the sensitivity to aliasing errors, and how one locally averages the results to be meaningful for grid-adaptation. The method is mainly assessed in several channel flows where the behavior of LES is relatively well known and where we therefore can judge the optimality of the resulting grids. The method is finally applied to the transonic flow around an airfoil to verify that it works for more realistic and complex problems.

\section{Methodology}
Any grid-adaptation method necessarily starts by estimating the spatial distribution of the error generation process (\ie, the residual, or source term for the solution error). It then proceeds by generating a new mesh that would minimize the residual field. The main focus of this paper is on the first step, specifically on how to compute the estimated residual field in the context of a Discontinuous Galerkin Spectral Element Method (DGSEM) code for grids with primarily hexahedral elements.

\subsection{The DGSEM code FLEXI\label{numerics}}
The simulations are run using the Discontinuous Galerkin Spectral Element Method (DGSEM) framework FLEXI. The code is developed in the Numerics Research Group at the University of Stuttgart \cite{Krais2021}. It is open source and can be downloaded from \url{http://www.flexi-project.org/}. The DGSEM \cite{Krais2021,hindenlang2012,Kopriva2009} uses piecewise smooth functions in every element, but allows for discontinuities at element interfaces. However, the residual estimation procedure (\sref{errorest}) and subsequent grid-adaptation can be transferred to any other DG or related scheme with minimal changes, as long as an element-local filter operation can be specified. Thus, the details regarding implementation given below serve to elucidate the procedure for this specific DG variant.

The computational domain $\Omega$ is partitioned in non-overlapping elements $C\in\Omega$. The solution in each element is described using a polynomial basis of degree $P$. We discretize the compressible Navier-Stokes equation using the same polynomial basis function as ansatz function in each three-dimensional element. The polynomials in each direction are represented using nodal 1D Lagrange basis functions on $P+1$ Legendre-Gauss-Lobatto points. Other node choices are possible, here we stick with these to be able to formulate the so-called split forms for the advective operators. To apply common integration rules and to build an efficient scheme, we define the three-dimensional hexahedral reference element $[-1, 1]^3$. We create the three-dimensional basis as the tensor product of three one-dimensional polynomials. We integrate the Navier-Stokes equation in space using the associated $P+1$ Legendre-Gauss-Lobatto quadrature to compute the projection integral. This results in the very efficient DGSEM. At the element boundaries we use a Riemann solver \eg \cite{Toro2009}. To take the parabolic terms into account, the BR1 lifting procedure \cite{Bassi1997} is applied. According to the method of lines we advance the resulting ODE in time by applying an explicit-in-time low storage Runge-Kutta method with optimized stability region \cite{Niegemann2012}.

We rely on skew-symmetric splitting of the advective terms of the compressible Navier-Stokes equations for stability \cite{Flad2017}, since in implicitly filtered LES the simulations are typically under-resolved. To obtain a dissipation-free and kinetic-energy preserving semi-discrete system we use the skew-symmetric split fluxes by Pirozzoli \cite{Pirozzzoli2010}. The Roe numerical flux is used to solve the Riemann problem at the cell interfaces \cite{Roe1981}. We use the Vreman model for explicit sub grid scale modeling in the following simulation \cite{vreman:04}.

\subsection{Estimating the LES Residual}\label{errorest}

Consider a general evolution equation $\partial q / \partial t = R(q)$ solved numerically on a grid with spacing $\Delta$,
the solution of which can be denoted by
\begin{equation} \label{eqn:N_definition}
\mathcal{N}_\Delta (q_\Delta) \equiv R_\Delta(q_\Delta) - \frac{\partial q_\Delta}{\partial t} = 0
\,,
\end{equation}
where $R_\Delta$ implies that $R$ is approximated at grid-spacing $\Delta$
and $q_\Delta$ means the solution to the discrete problem at this grid-spacing.
The error in this equation is then $q_\Delta - q$, which satisfies the error equation
\begin{equation} \nonumber
\frac{\partial \mathcal{N}_\Delta}{\partial q}
\left( q_\Delta - q \right)
\approx
\underbrace{\mathcal{N}_\Delta (q_\Delta)}_{=0}
-
\underbrace{\mathcal{N}_\Delta (q)}_{\mathcal{F}}
\,,
\end{equation}
where $\mathcal{F}$ is the residual, \ie, the source term for error.
The residual is defined based on the exact solution $q$ which is, of course, not known.
All grid-adaptation methods therefore involve some type of process for estimating the residual from the numerical solution $q_\Delta$, for example using the leading terms in Taylor expansions of the numerical operators or by directly approximating $q$ by interpolating $q_\Delta$ onto a refined mesh.
Neither approach works in large eddy simulation (LES) which by definition seeks a solution to the coarse-grained Navier-Stokes equation.
While one could interpolate the solution $q_\Delta$ onto a finer grid, in reality the solution on this finer grid should have developed smaller scales due to the broadband nature of turbulence; the resulting residual estimate would therefore be incorrect.
For the same reason, the fact that an LES solution is, by definition, a rough solution far from the asymptotic range of numerical convergence means that many terms in a Taylor expansion should be expected to be large, and thus one could not approximate the error behavior from the leading term only as is traditionally done in numerical analysis.

Toosi and Larsson \cite{toosi:20,toosi:21} proposed that the modeling residual (\ie, the residual due to imperfect resolution/modeling of the small scale turbulence, excluding the residual due to numerical errors)
in LES can be estimated in a post-processing step using low-pass test-filtering.
Specifically, they argued that the residual must be estimated at an imagined coarser resolution $\tf{\Delta}$ than the resolution $\Delta$ used in the actual LES.
Their argument went as follows:
Assume that the LES equations at resolution $\Delta$ (\ie, the LES solved in the code) can be written as
in \eref{eqn:N_definition}.
The residual at the test-filtered (or additionally coarse-grained) level is then
$\mathcal{N}_{\tf{\Delta}} (\tf{q})$,
where $\tf{q}$ is the exact solution restricted (or test-filtered) to the resolution $\tf{\Delta}$.
Note that one could not define the residual with $q$ directly, since this would contain smaller scales that would then be double-counted due to the possible subgrid-model in the LES.
Toosi and Larsson \cite{toosi:20,toosi:21}
then suggested that $\tf{q}$ can be approximated by $\tf{q_\Delta}$, and thus defined the approximate modeling residual as
\begin{equation} \nonumber
\mathcal{F}_{\tf{\Delta}} = 
\mathcal{N}_{\tf{\Delta}} \left( \tf{q_\Delta} \right)
= R_{\tf{\Delta}} \left( \tf{q_\Delta} \right) - \frac{\partial \tf{q_\Delta}}{\partial t}
\,.
\end{equation}
Assuming that the test-filter commutes with the time-derivative and using \eref{eqn:N_definition} then yields
\begin{equation} \label{eqn:F_as_rhs_evaluations}
\mathcal{F}_{\tf{\Delta}}
= R_{\tf{\Delta}} \left( \tf{q_\Delta} \right) - \reallywidehat{ R_\Delta \! \left( q_\Delta \right) }
\,.
\end{equation}

In the context of subsonic flows without shocks or other multi-physics effects (heat release, multi-phase, etc), 
the modeling residual for the momentum equation then becomes
\begin{align} \label{eqn:F_for_momentum}
\mathcal{F}_{i,\tf{\Delta}} &
= \reallywidehat{ 
\frac{\partial}{\partial x_j} \left[
\rho u_i u_j + p \delta_{ij} + \tau_{ij}^{\rm mod} (\rho, u_i, \Delta) - \sigma_{ij}
\right] } \\ \nonumber
& -
\frac{\partial}{\partial x_j} \left[
\tf{\rho} \tf{u}_i \tf{u}_j + \tf{p} \delta_{ij} + \tau_{ij}^{\rm mod} (\tf{\rho}, \tf{u}_i, \tf{\Delta} ) - \tf{\sigma_{ij}}
\right]
\,,
\end{align}
where the test-filtered velocity should be understood as a Favre-weighted filter.

All operations above were defined based on the instantaneous flow fields, and thus the
estimated modeling residual $\mathcal{F}_{i,\tf{\Delta}}$ is necessarily a chaotic field in space and time.
Assuming that one seeks a stationary grid (\ie, that one runs a full LES and then adapts the grid before running another full LES),
one must then reduce the residual field in time (and possibly in any spatially homogeneous directions).
While there is no theoretical reason for any specific choice of reduction, Toosi and Larsson~\cite{toosi:20, toosi:21} suggested using the L2 norm.
In addition, they argued that the test-filter should be chosen as a uni-directional one, providing filtering in one direction at a time.
For hexahedral elements, this implies that one should compute 3 different residual fields (using test-filtering in each of the 3 directions of a hexahedral element) which then provides information about the lack of resolution in each direction separately; this then allows for anisotropic grid-adaptation.

Putting these things together, the final time-averaged modeling residual is written as
\begin{equation} \label{eqn:FtoG}
G(\vec{x},\vec{n}) = \sqrt{\langle \mathcal{F}_i(\vec{x},\vec{n})  \mathcal{F}_i (\vec{x},\vec{n}) \rangle }
\,,
\end{equation}
where $\vec{x}$ denotes the spatial location,
$\vec{n}$ denotes the direction of the uni-directional test-filter,
and the angular brackets denote averaging in time, spatially homogeneous directions, and possibly locally in space as well (to be discussed below).

We emphasize that the subgrid model contribution in the residual~(\ref{eqn:F_for_momentum}) must be computed with the correct length scale: when evaluated for the test-filtered field, it must use the length scale of the imagined $\tf{\Delta}$ resolution. This length scale must be estimated from the properties of the test filter.

For implicit LES run without an explicit subgrid model, the subgrid model terms are simply zero in the residual estimate.

\subsection{Computation of the LES modeling residual in a DGSEM code}

The implementation of the LES residual estimator described above in a
Discontinuous Galerkin Spectral Element Method (DGSEM)
requires some care and specific implementation details, which are described in this section.

The low-pass test-filtering operator \makebox[\widthof{$Q$}]{$\widehat{\cdot}$} is realized as a uni-directional modal cut-off filter, which removes the highest modes of the polynomial representation in one direction only.
Throughout this work we used \nth{5} order polynomials, \ie $P=5$ where $P$ represents the polynomial order, to represent the solution in each direction resulting in a \nth{6} order accurate scheme, and chose to low-pass filter by setting the coefficients of the \nth{4} and \nth{5} order polynomials to zero.
Importantly, this was done in only one direction, with the high-order polynomial coefficients in the other directions remaining unchanged.
A nice feature of the DGSEM method is that the filtering in one element is independent from all other elements, and thus it works equally well in an unstructured grid (of hexahedral elements).

The LES residual $\mathcal{F}_{\tf{\Delta}}$ could be computed using either the generic \eref{eqn:F_as_rhs_evaluations}
or the more flow-specific \eref{eqn:F_for_momentum}.
The most convenient and easy-to-implement option is the former one, as most codes (especially ones with explicit time-stepping) have functions to compute $R_\Delta$ for a given solution field.
Given an instantaneous LES field $q_\Delta$, it is then trivial to compute $R_\Delta(q_\Delta)$ using a single function-call in the code.
The resulting field is then low-pass test-filtered three times, one for each natural direction of each hexahedral element, to form three different instantiations of the second part of \eref{eqn:F_as_rhs_evaluations}.

In a separate process, the LES solution itself is test-filtered to form $\tf{q_\Delta}$.
In the present work, we apply the test-filter to the conserved variables which then implies that the filtered velocity is Favre-weighted.
This test-filtered LES solution is then fed into the computation of $R_{\tf{\Delta}}(\tf{q_\Delta})$ to complete \eref{eqn:F_as_rhs_evaluations}.
This last step requires the most care, as the function in the LES code will (for the given grid and polynomial order) compute $R_\Delta(\tf{q_\Delta})$ rather than the correct $R_{\tf{\Delta}}(\tf{q_\Delta})$.
If we are interested only in the LES modeling residual rather than the numerical errors (as was the case in Toosi and Larsson~\cite{toosi:20, toosi:21}), then the only modification required is to ensure that the length scale in the subgrid model is reflective of the coarse-grained state.
This is the option chosen in this work.
In principle one could also use a lower polynomial order when evaluating $R_{\tf{\Delta}}(\tf{q_\Delta})$ as a way to also approximately account for the numerical error, but this is left for future work.

The present simulations use the Vreman subgrid model~\cite{vreman:04} but in a modified form in which the length scale is taken as
\begin{equation} \nonumber
\Delta = \dfrac{\mathcal{V}^{1/3}}{P+1}
\,,
\end{equation} 
where $\mathcal{V}$ is the element volume and $P$ is the polynomial order, taken as $P=5$ in this work.
Since we use a uni-directional filter that keeps modes up to the $M$th order (taken as $M=3$ in this work),
the consistent length scale after test-filtering is
\begin{equation} \nonumber
\tf{\Delta} = \left( \frac{P+1}{M+1} \right)^{1/3} \, \Delta
\,.
\end{equation}
Since the model implemented in the FLEXI code uses an isotropic length scale, this implies that the eddy viscosity at the test-filtered level is simply a factor of $[(P+1)/(M+1)]^{2/3}\approx 1.59$ larger than the eddy viscosity computed by the code.

The norm operation in equation \eref{eqn:FtoG} can be tricky because the solution is represented as a polynomial in an appropriate number of collocation points, but the included squaring operator in the norm operation needs a higher order of polynomial, thus more collocation points. The approach we took is to map the $\mathcal{F}_{\tf{\Delta}}$ from the original $P$ degree of polynomial to $2P$ degree of polynomial and its associated collocation points, and then take the square of the polynomial at that order. This way we got an exact representation of the term $\mathcal{F}_i(\vec{x},\vec{n})  \mathcal{F}_i (\vec{x},\vec{n})$.

Another issue has risen in the computation, which is the cell edges. Since the DG solution is discontinuous at the element boundaries and local projections tend to produce large gradients at these boundaries, the local gradients at the cell boundaries in an underresolved setting can overshoot. This basically results in the residual estimator showing the element boundaries as high error regions, when this is in fact a numerical effect that is excluded from its goal, to find high error regions in the modeling of turbulence. To mitigate this issue, and to stay true to the cell based nature of the discontinuous Galerkin method, we average the $\mathcal{F}_i(\vec{x},\vec{n}) \mathcal{F}_i(\vec{x},\vec{n})$ term over the whole cell before the periodic direction or time averaging. This averaging is also included in the $\langle \rangle$ averaging operator that has been used so far.

\subsection{Finding the optimal grid-spacing}
Once the estimated residuals for test-filtering in each direction are computed, the optimal element size in each direction can be found from an optimization problem and an assumed model for how the residuals vary under grid-refinement.

Following Toosi and Larsson~\cite{toosi:20}, the directional residual $G(\vec{x}, \vec{n})$ is assumed to vary as
\begin{equation}  \label{eqn:G_to_g}
G(\vec{x}, \vec{n}) = g(\vec{x}, \vec{n}) \, \Delta (\vec{x}, \vec{n})^\alpha
\,.
\end{equation}
In the asymptotic limit of convergence, the power $\alpha$ would be the order-of-accuracy of the numerical method.
In the context of LES which is, by definition, far from the asymptotic limit of convergence, the power $\alpha$ is instead related to the spectral behavior of turbulence near the grid cut-off. Since the spectral slope of inertial range turbulence varies across flow types and directions, it is clear that $\alpha$ should in theory also vary analogously.
The more important point is that the value of $\alpha$ comes entirely from turbulence physics and not from numerical analysis.
Rather than try to find the ``correct'' $\alpha$ field, we follow the suggestion of Toosi and Larsson~\cite{toosi:17} and simply assume a constant value of $\alpha=2$ throughout the flow domain.
This is almost certainly larger than the ``correct'' value in most flow scenarios; erring on the side of a larger value produces lower variations in the optimal grid-spacing, which can be beneficial in practice.
With this model for the residual scaling, the ``residual density'' $g(\vec{x}, \vec{n})$ can be computed by evaluating \eref{eqn:G_to_g} for the residual estimated on the current grid, in each direction.

We then define the cost functional
\begin{equation} \nonumber
\mathcal{J}[\Delta
(\vec{x},\vec{n})
] = \iiint\limits_{V} \left[ \left(\sum \limits_{m} g(\vec{x},\vec{n}_m)^\beta 	\Delta(\vec{x},\vec{n}_m)^{\beta\alpha}\right) ^{1/\beta}
+ \dfrac{\lambda}{\prod \limits_{m} \Delta(\vec{x},\vec{n}_m) }\right] dV
\end{equation}
where the first term accounts for the sum of residuals in all directions in a $\beta$-norm (note that all factors are positive)
and the second term accounts for the computational complexity (defined here as the number of elements) of the new grid,
with $\lambda$ being a Lagrange multiplier.
The solution to this particular calculus-of-variations problem is that the variation of the integrand with respect to the
$\Delta(\vec{x},\vec{n}_i)$ field is zero for each hexahedral direction $i=1,2,3$.
Using the abbreviated notation $\Delta_i = \Delta(\vec{x},\vec{n}_i)$ and $g_i = g(\vec{x},\vec{n}_i)$, this is
\begin{equation} \nonumber
\alpha g_i^\beta \Delta_{i, {\rm opt}}^{\beta \alpha -1}
\left( \sum \limits_{m} g_m^\beta \Delta^{\beta\alpha}_{m, {\rm opt}} \right)^{1/\beta-1}
- \dfrac{\lambda}{\Delta_{i, {\rm opt}} \prod \limits_{m}\Delta_{m,{\rm opt}}} = 0
\ , \ \ 
i=1,2,3
\,,
\end{equation}
where the subscript ``opt'' has been added to indicate that this is the $\Delta(\vec{x},\vec{n})$ field that minimizes the cost functional $\mathcal{J}$.
Some rearrangement yields
\begin{equation} \nonumber
\left( g_i \Delta_{i, {\rm opt}}^\alpha \right)^\beta
= \dfrac{\lambda/\alpha}{\prod \limits_{m}\Delta_{m,{\rm opt}}}
\left( \sum \limits_{m} g_m^\beta \Delta^{\beta\alpha}_{m, {\rm opt}} \right)^{1-1/\beta}
\ , \ \ 
i=1,2,3
\,.
\end{equation}
This is valid for all $i=1,2,3$ but the right-hand-side is the same for all $i$; thus
the optimal residual $G_{i,{\rm opt}} = g_{i, {\rm opt}} \Delta_{i, {\rm opt}}^\alpha$ is the same in all three directions.
This can then be exploited to find that
\begin{equation} \label{eqn:optimal_gridspacing}
g_{i, {\rm opt}} \Delta_{i, {\rm opt}}^\alpha \, 
\underbrace{\prod \limits_{m}\Delta_{m,{\rm opt}}
}_{\mathcal{V}_{\rm opt}}
= \Lambda
\ , \ \ 
i=1,2,3
\,,
\end{equation}
where $\Lambda$ is a re-defined Lagrange multiplier
and $\mathcal{V}_{\rm opt}$ is the volume of the optimal element;
this equation shows that the element-integrated residual should be equi-distributed in both space and direction.

The optimal grid-spacing is then found by using a root-finding procedure to find the value of the modified Lagrange multiplier $\Lambda$ for which the computational complexity (=the number of elements) is as desired.
In each step, \eref{eqn:optimal_gridspacing} is used to find the optimal $\Delta(\vec{x},\vec{n})$ field.
The process was documented originally in Toosi and Larsson \cite{toosi:20} and is provided here for completeness.

We note that the final result (\eref{eqn:optimal_gridspacing}) is independent of the value of $\beta$, and thus the choice of norm in the cost functional is immaterial.

\subsection{Overall process}

The grid-adaptation process is implemented entirely as a post-processing approach that is performed between LES runs, with no adaptation occurring on-the-fly during a run.
After a simulation, the residual field is computed from multiple instantaneous snapshots of the solution, with averaging in time and possibly suitable spatial directions.
We then compute the residual ``density'' field $g(\vec{x}, \vec{n})$ from the residual scaling model, and the use iterative root-finding on the Lagrange multiplier $\Lambda$ to find the optimal grid-spacing field $\Delta(\vec{x}, \vec{n})$ with the desired computational complexity.
From this point we have multiple choices for the actual creation of the adapted grid.
One approach is to modify the existing grid, most simply by splitting individual elements in those directions for which the element-integrated directional residual exceeds some threshold.
A second choice is to re-generate the grid to more closely follow the optimal target.
This latter approach has mainly been followed for simplex elements in the literature~\cite{Park:2016,Park:2018}.
Since we desire meshes with hexahedral cells in this work, we resort to human intervention in the grid-generation process: we build the grids manually, but aimed to mimic the optimal grid-spacing field as closely as possible.
In future work we hope to automate this aspect.

We note that the estimated residuals do not say anything about whether the grid is sufficiently fine to be considered converged; instead, any judgment about convergence must be made for specific quantities-of-interest (\eg, mean profiles, drag or lift coefficients, etc).
This judgment must be made by the user, prior to deciding whether to create an adapted grid.

\section{Application to channel flows}\label{results}
In this and the following section, we use the abbreviated notation
$\Delta_i = \Delta(\vec{x}, \vec{n}_i)$ where $\vec{n}_i$ is a unit vector in one of the three natural directions of a hexahedral element.

As the first step the algorithm is applied to the canonical test case of the plane turbulent channel flow. 
The appeal of channel flow is that it is one of very few problems where the ``optimal'' grid is known to some degree; it is then a good test of a grid-adaptation method to see whether it can arrive at something close to the known ``optimum''.
The domain is taken as $(10h,2h,3h)$ where $h$ is the channel half-height.
Two different Reynolds numbers of $Re_\tau=550$ and $Re_\tau=\num{2000}$ are targeted, with the lower used to compare the adapted grids to those by Toosi and Larsson~\cite{toosi:20} and the higher used to test the grid-adaptation process when a wall-model is used and when mortar elements are used in the adapted grid.

\subsection{Channel flow at $Re_\tau = 550$, validation of the original results}
The point of this case is to reproduce the original results of \citep{toosi:20}, establishing that the discontinuous Galerkin method is suitable for the grid-adaptation process this way. We start with a uniform mesh of 2 cells along the channel half-width $h$, 10 cells along the streamwise direction and 3 cells along the spanwise direction with each cell containing a basis of \nth{5} order polynomials. This is a similar setup to \citep{toosi:20} (although a tad ``finer''), where they start with grid spacings $(0.2h,0.1h,0.2h)$. 

For each iteration, the grid-adaptation algorithm is used as described, with each new grid having 4 times the previous number of elements.
It is constructed such that it is smoothly stretched along the wall-normal direction, agreeing with the near wall $\Delta_2$ and the center line $\Delta_2$, and has uniform cells along the streamwise $x$ and spanwise $z$ directions with the minimum suggested $\Delta_1$  and $\Delta_3$, respectively.
The wall-normal stretching is realized using a bell shaped 
\begin{equation}
DR(s) = 1+\left(\dfrac{\Delta_{2,\text{min}}}{\Delta_{2,\text{max}}}-2\right)\cdot\dfrac{e^{-(s\cdot f)^2}-e^{-f^2}}{1-e^{-f^2}}
\end{equation}
element distribution with $s\in[-1,1]$. The ration $\frac{\Delta_{2,\text{min}}}{\Delta_{2,\text{max}}}$ describes the difference between the smallest and the largest element in $y$ direction, $f$ denotes the scaling factor and the result $DR$ defines the element local weight which is later normalized to get the element size distribution in $s\in[-1,1]$.
To ensure compatibility with Toosi and Larsson \citep{toosi:20}, we use the Vreman subgrid scale model for LES closure \citep{vreman:04} with the same model constant 0.03.
The grid-adaptation process is terminated when both the mean velocity profile and the Reynolds stresses change by at most a few percent between adaptive iterations.

\begin{figure}[!t]
\centering
\includegraphics{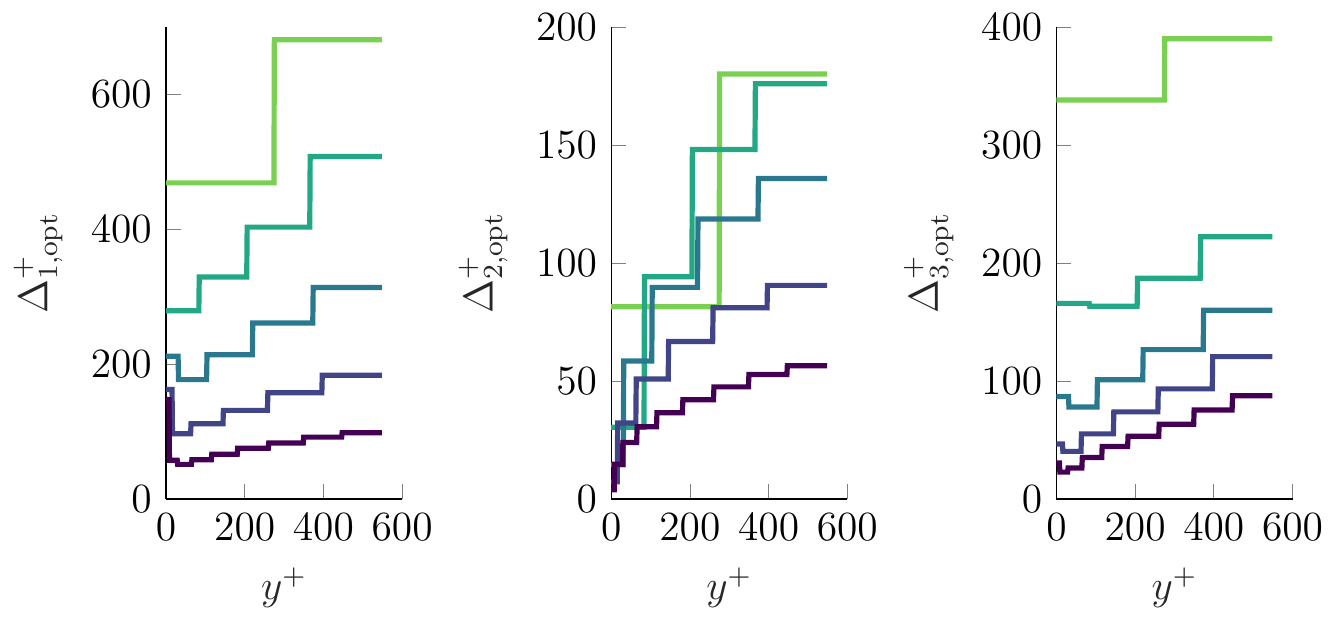}
\includegraphics{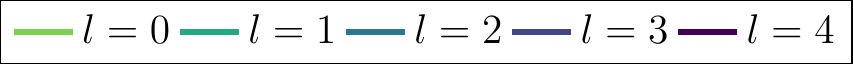}	
\caption{Optimal element sizes for the channel with $Re_\tau=550$ after adaptive iteration $l$.}
\label{fig:DeltaX_WRLES}
\end{figure}

The resulting grid-spacing distributions after each adaptive iteration are shown in \fref{fig:DeltaX_WRLES},
with some quantities listed in \tref{tab:WRLES}.
The staircase pattern in \fref{fig:DeltaX_WRLES} corresponds to individual elements; recall that the residual is averaged over each element, and thus the optimal grid-spacing field inherits this step-wise nature.

\begin{table}[]
\caption{Key properties of the $Re_\tau=550$ channel flow case for different adaptive iterations $l$.
}
\label{tab:WRLES}
\centering
\resizebox{\columnwidth}{!}{\begin{tabular}{|l|ccc|ccc|ccc|c|c|c|}
\hline
Iteration &
\multicolumn{3}{c|}{$N_{i}$} &
\multicolumn{3}{c|}{$\Delta_{i}(y=h)/h$} &
\multicolumn{3}{c|}{$\Delta^+_{i}(y=0)/(P+1)$} &
\texttt{\#Elems} &
\texttt{\#DOF} &
$Re_\tau$ \\ \hline
$l=0$ & 10 & 2 & 3  & 1.00 & 0.50 & 1.00  & 91.67 & 45.83 & 91.67 & 60    & 12960   & 178.50 \\
$l=1$ & 12 & 4 & 5  & 0.83 & 0.25 & 0.60  & 76.39 & 13.93 & 55.00 & 240   & 51840   & 547.80 \\
$l=2$ & 20 & 5 & 10 & 0.50 & 0.20 & 0.30  & 45.83 & 5.13  & 27.50 & 1000  & 216000  & 498.11 \\
$l=3$ & 31 & 6 & 21 & 0.32 & 0.17 & 0.14  & 29.57 & 2.57  & 13.10 & 3906  & 843696  & 511.73 \\
$l=4$ & 50 & 9 & 36 & 0.20 & 0.11 & 0.08  & 18.33 & 1.28  & 7.64  & 16200 & 3499200 & 556.57 \\ \hline
\end{tabular}}
\end{table}

The sequence of grid-spacings found are similar to those by Toosi and Larsson \citep{toosi:20} using the same overall algorithm and residual estimation method but a totally different code (a finite-difference code).
Table~\ref{tab:WRLES} indicates that we are sufficiently resolved after iteration $l=4$, with $\Delta(y=0,\vec{n}_i)^+ \approx (18,1,8)$ being viewed as an acceptable (albeit a bit fine) grid by the LES community.
The computed wall shear stress, here shared as $Re_\tau$, converges to the correct value. 

An interesting observation in \fref{fig:DeltaX_WRLES} is that the suggested grid-spacings along $x$ and $z$ are higher at the viscous sublayer than in the buffer layer.
This is actually the correct behavior since the wall-parallel length scales in the viscous sublayer become larger than in the buffer layer
(\eg, Jimenez~\cite{jimenez:12}), and it is interesting to see that the residual estimator correctly reflects this.

\begin{figure}[!t]
\centering
\includegraphics{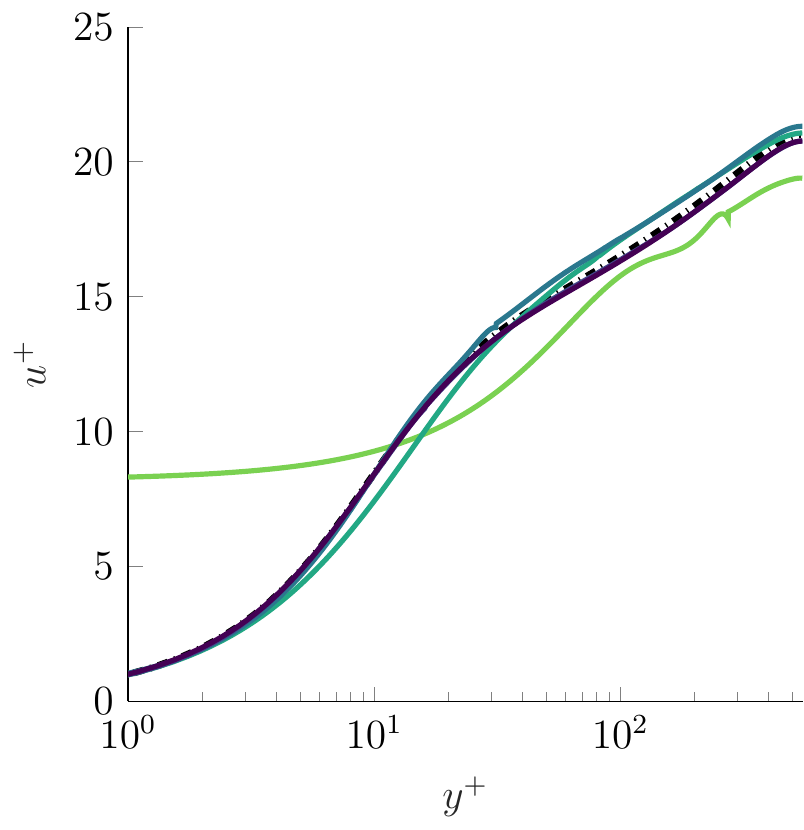}
\includegraphics{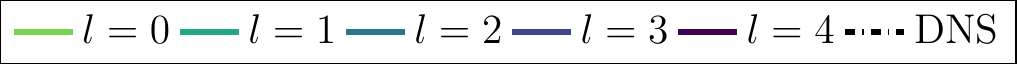}
\caption{Mean velocity profiles for the $Re_\tau=550$ channel for different adaptive iterations $l$.}
\label{fig:umean_WRLES}
\end{figure}

\begin{figure}[!t]
\centering
\includegraphics{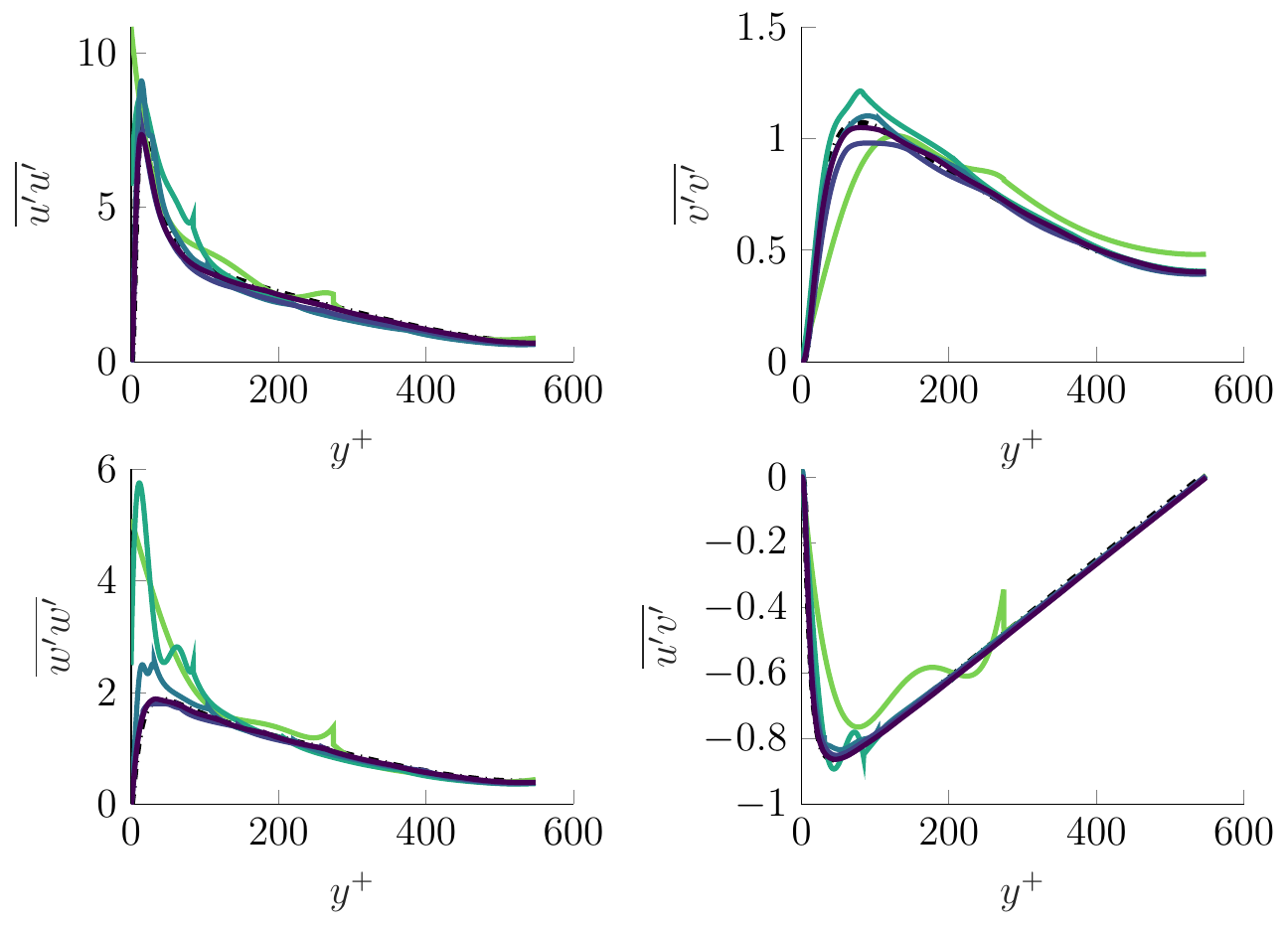}
\includegraphics{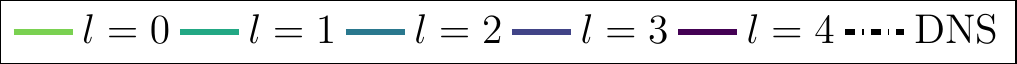}
\caption{Reynolds stresses for the $Re_\tau=550$ channel for different iteration $l$.}
\label{fig:reynolds_stresses_WRLES}
\end{figure}

The mean velocity profiles in inner units are shown in \fref{fig:umean_WRLES}, and show good agreement and convergence towards the DNS data of Lee and Moser \cite{lee:15:channel} at $l=4$. The Reynolds stresses, shown in \fref{fig:reynolds_stresses_WRLES}, also have good convergence on the DNS data. Different from the mean velocity profiles though, the Reynolds stresses have already converged by $l=3$. For the first iteration $l=0$, we can see the unsteady characteristics of the discontinuous Galerkin scheme, that the element boundaries are clearly visible for $\overline{u'v'}$. Additionally, the flow field is severely underresolved and we observe the typical overshoots of DG schemes in these cases at the element boundaries as well as the inability to fulfill the weakly enforced no-slip condition. Both behaviors are to be expected and are a tell-tale sign of insufficient resolution. These oscillations in the Reynolds stresses, however, are mitigated at higher resolutions and therefore convergence towards the ``smooth'' reference DNS solution is obtained. The overall results for this case show that this grid-adaptation algorithm is applicable to the discontinuous Galerkin framework, and allow us to move further with configurations and flows.

\subsection{Channel flow at $Re_\tau = \num{2000}$, application with meshes containing mortar elements}

\begin{figure}[!t]
\centering
\includegraphics[width=0.75\columnwidth]{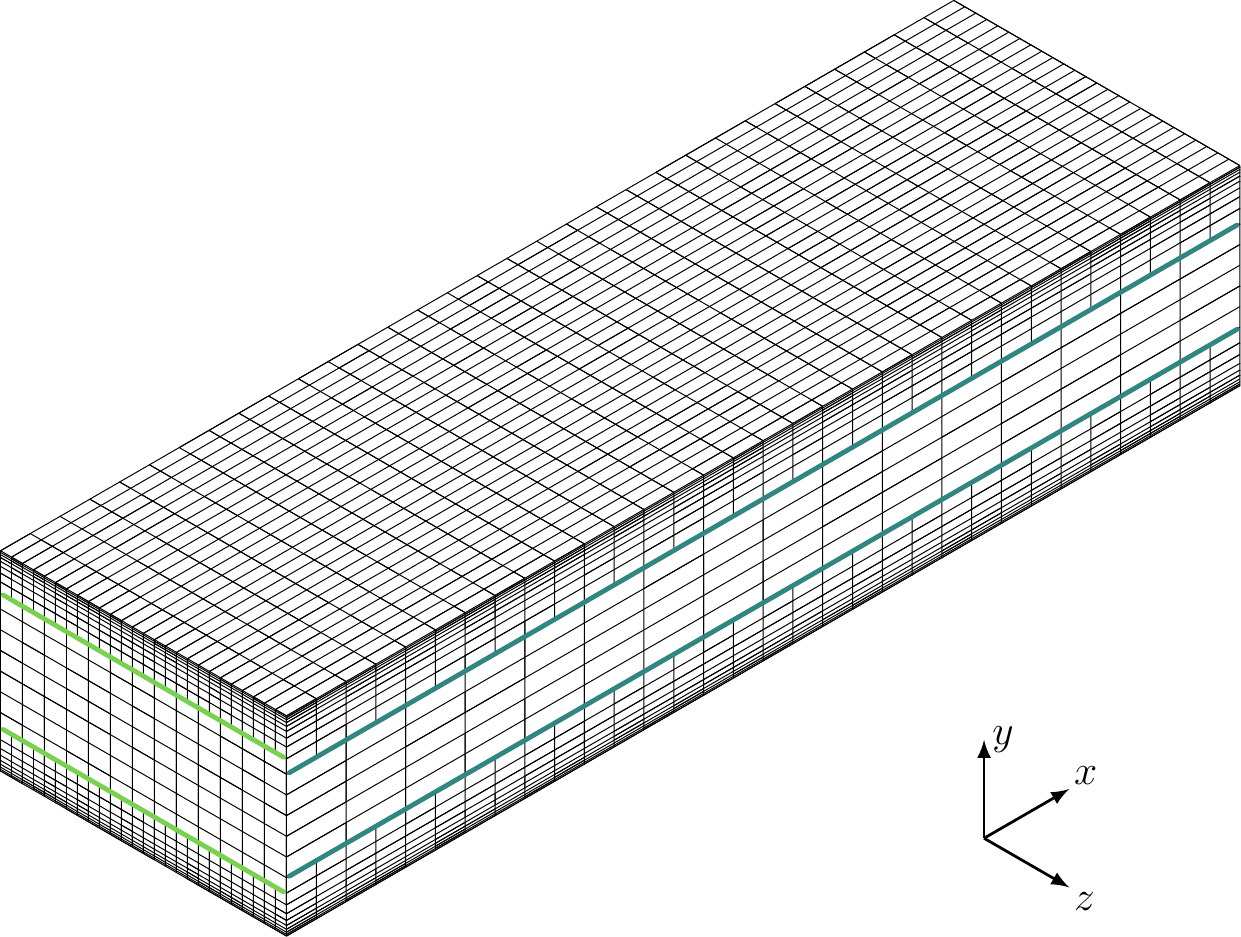}
\caption{An example of a mortared mesh, with mortars ``along'' different directions marked with different colors, teal for $x$ direction and light green for $z$ direction.}
\label{fig:image_Mortar}
\end{figure}

Next, we test the full unstructured mesh ability of the discontinuous Galerkin method in conjunction with the grid-adaptation using mortar elements. Mortar elements are elements whose faces do not match up one-to-one with their neighbors, instead a single face of an element may have more than one neighbor. A representation of this kind of a mesh can be seen in Figure \ref{fig:image_Mortar}, showing mortars ``along'' $x$ (teal line) and $z$ (light green line) directions. Note that these are actually planes with $y$ direction as normal vectors, shown here as lines to increase the clarity of the representation. The advantage of these elements is that they allow for substantial cost savings compared to fully structured grids for flows with massively different length scales.
The benefit of mortars comes when the optimal wall-parallel grid-spacing changes by large amounts across the channel.
We therefore increase the Reynolds number to $Re_\tau = \num{2000}$ to get a larger ratio of length scales.

\begin{figure}[!t]
\centering
\includegraphics{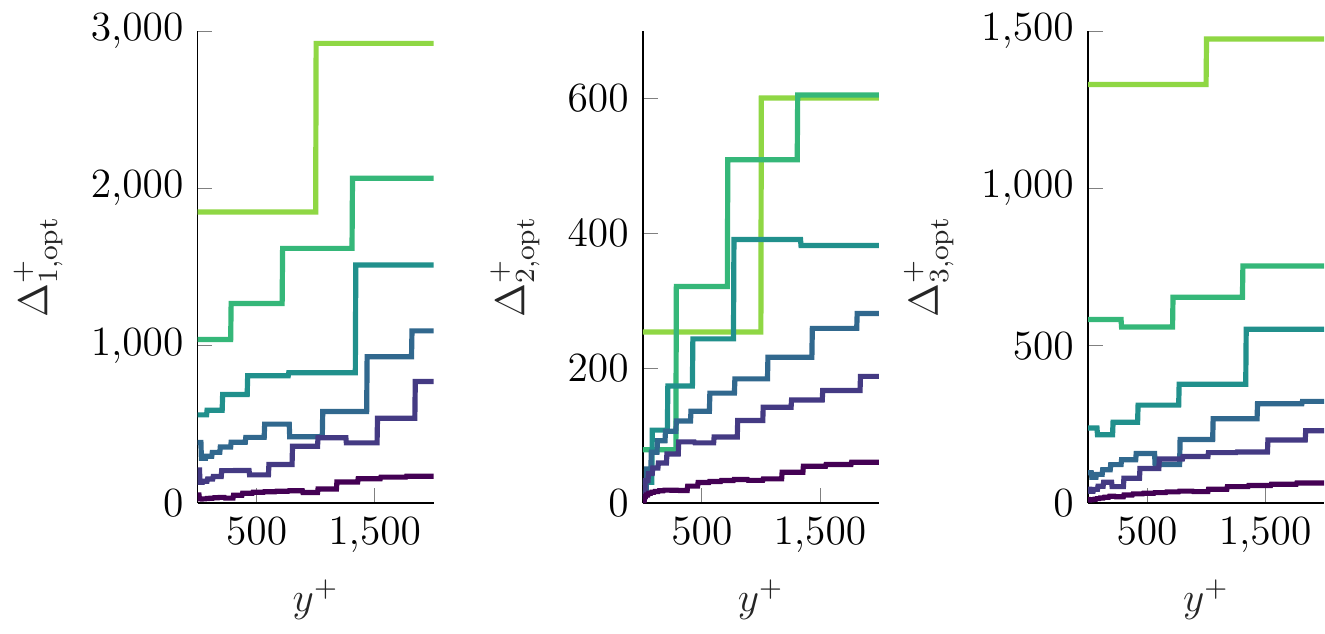}
\includegraphics{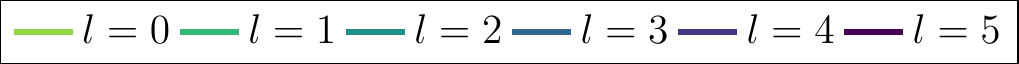}
\caption{Optimal gridspacing according to $G$ for the $Re_\tau=\num{2000}$ channel for different iteration $l$.}
\label{fig:DeltaX_Mortar}
\end{figure}

\begin{figure}[!t]
\centering
\begin{subfigure}[b]{0.64\textwidth}
\centering
\includegraphics{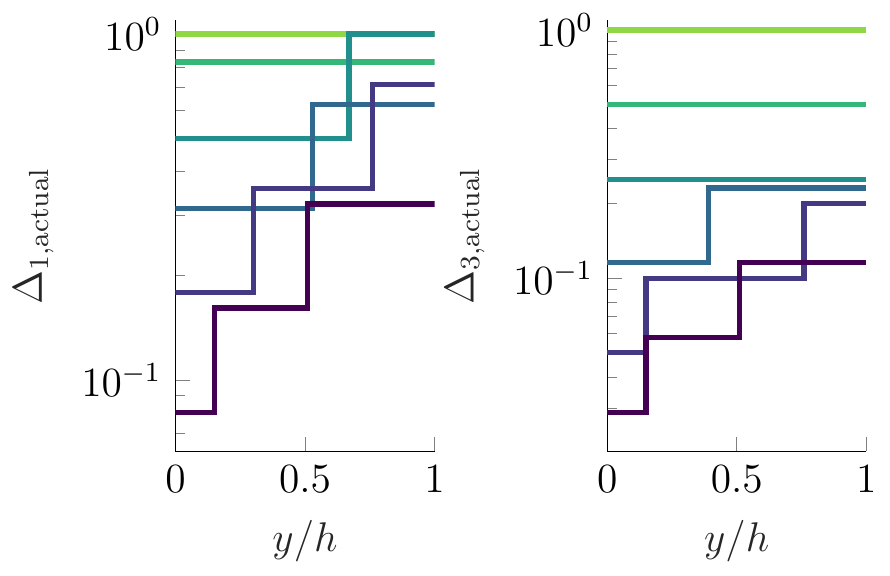}
\includegraphics{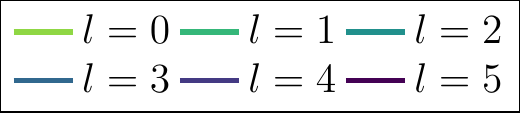}
\caption{The actual grid spacings for the streamwise and spanwise direction mortar elements of the $Re_\tau=\num{2000}$ channel, non-dimensionalized by the largest cell along respective direction.}
\end{subfigure}
\begin{subfigure}[b]{0.32\textwidth}
\centering
\includegraphics{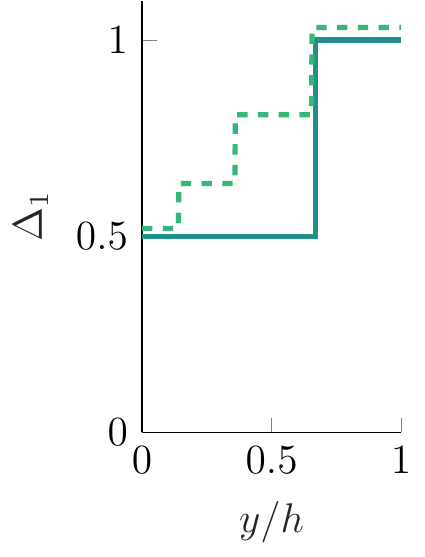}
\includegraphics{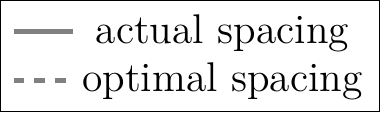}
\caption{Comparison of actual and optimal grid-spacing for $l=2$ mesh using $l=1$ as reference.}
\end{subfigure}
\caption{Overview over the actual and optimal gridspacing for the channel containing mortar elements.}
\label{fig:mortar_meshes}
\end{figure}

The output of the grid-adaptation optimization problem $\Delta_{\rm opt}(\vec{x}, \vec{n})$ is shown in \fref{fig:DeltaX_Mortar} after every adaptive iteration.
The mortar implementation in FLEXI is limited to 2-to-1 interfaces, and therefore we decide where to place the mortar interfaces as follows.
We first find the smallest required element spacing in a direction (generally very close to the wall).
We then maintain this element spacing until we reach a wall-distance at which the optimal grid-spacing has at least doubled: we then insert a 2-to-1 mortar interface at that wall-distance, and continue.
This results in the actual grid-spacing distribution in \fref{fig:mortar_meshes}. 
We note that the mortar interfaces can (and actually do) occur at different wall-distances for the $x$- and $z$-directions.
Finally, the grid is stretched in the $y$-direction in the same way as for the $Re_\tau=550$ case.

We use the same domain size and numerical parameters for the $Re_\tau=\num{2000}$ test case as already introduced in the $Re_\tau=550$ case. The suggested element sizes are again visualized in \fref{fig:DeltaX_Mortar} and the resulting meshes are described in \fref{fig:mortar_meshes}. It clearly indicates that starting at $l=2$, we can use mortar interfaces in both $x$ and $z$ because the suggested grid spacing starts to increase by a factor of more than two compared to the minimum grid spacing in this direction. For $l=2$ the interfaces in $x$ and $z$ are located at the same wall-normal distance. Thus, the resulting mesh $l=3$ will contain one mortar interface. At $l=3$ these locations are shifted and are no longer identical as indicated in \fref{fig:image_Mortar}. For $l=4$ we even get a second mortar interface and therefore have a largest length ratio of four along the wall-parallel directions between the wall elements and the elements in the channel center. The grid for $l=3$ is visualized in \fref{fig:image_Mortar} showing the different mortar locations in the $z$ and $x$ plane.

\begin{table}[]
\caption{Number of elements for the $Re_\tau=\num{2000}$ channel with mortar meshes.
}
\label{tab:Mortar}
\centering
\resizebox{!}{0.10\columnwidth}{\begin{tabular}{|l|ccclcccccclcc|ccclcccccclcl|}
\hline
&
\multicolumn{13}{c|}{} &
\multicolumn{13}{c|}{} \\
\multirow{-2}{*}{Iteration} &
\multicolumn{6}{c}{\multirow{-2}{*}{$N_{1}$}} &				
\multicolumn{1}{c}{\multirow{-2}{*}{$N_{2}$}} &
\multicolumn{6}{c|}{\multirow{-2}{*}{$N_{3}$}} &
\multicolumn{6}{c}{\multirow{-2}{*}{$\Delta_{1}/h$}} &				
\multicolumn{1}{c}{\multirow{-2}{*}{$(\Delta_{2}/h)_\text{min}$}} &
\multicolumn{6}{c|}{\multirow{-2}{*}{$\Delta_{3}/h$}} \\ \hline	
$l=0$ &
\multicolumn{6}{c}{10} &
\multicolumn{1}{|c|}{2} &
\multicolumn{6}{c|}{3} &
\multicolumn{6}{c}{1.00} &
\multicolumn{1}{|c|}{0.50} &
\multicolumn{6}{c|}{1.00} \\
$l=1$ &
\multicolumn{6}{c}{12} &
\multicolumn{1}{|c|}{4} &
\multicolumn{6}{c|}{6} &
\multicolumn{6}{c}{0.83} &
\multicolumn{1}{|c|}{0.25} &
\multicolumn{6}{c|}{0.50} \\
$l=2$ &
&
{20} &
\multicolumn{2}{c}{} &
{10} &
&
\multicolumn{1}{|c|}{4} &
\multicolumn{6}{c|}{12} &
&
{0.50} &
\multicolumn{2}{c}{} &
{1.00} &
&
\multicolumn{1}{|c|}{0.17} &
\multicolumn{6}{c|}{0.25} \\
$l=3$ &
&
{32} &
\multicolumn{2}{c}{} &
{16} &
&
\multicolumn{1}{|c|}{11.5} &
&
{26} &
\multicolumn{2}{c}{} &
{13} &
&
&
{0.31} &
\multicolumn{2}{c}{} &
{0.63} &
&
\multicolumn{1}{|c|}{0.09} &
&
{0.12} &
\multicolumn{2}{c}{} &
{0.23} &
\\
$l=4$ &
\multicolumn{2}{c}{56} &
\multicolumn{2}{c}{28} &
\multicolumn{2}{c}{14} &
\multicolumn{1}{|c|}{13.5} &
\multicolumn{2}{c}{60} &
\multicolumn{2}{c}{30} &
\multicolumn{2}{c|}{15} &
\multicolumn{2}{c}{0.18} &
\multicolumn{2}{c}{0.36} &
\multicolumn{2}{c}{0.71} &
\multicolumn{1}{|c|}{0.07} &
\multicolumn{2}{c}{0.05} &
\multicolumn{2}{c}{0.10} &
\multicolumn{2}{c|}{0.20} \\
$l=5$ &
\multicolumn{2}{c}{124} &
\multicolumn{2}{c}{62} &
\multicolumn{2}{c}{31} &
\multicolumn{1}{|c|}{21} &
\multicolumn{2}{c}{104} &
\multicolumn{2}{c}{52} &
\multicolumn{2}{c|}{26} &
\multicolumn{2}{c}{0.08} &
\multicolumn{2}{c}{0.16} &
\multicolumn{2}{c}{0.32} &
\multicolumn{1}{|c|}{0.05} &
\multicolumn{2}{c}{0.03} &
\multicolumn{2}{c}{0.06} &
\multicolumn{2}{c|}{0.12} \\ \hline
\end{tabular}}
\end{table}	

\begin{table}[]
\caption{Grid-spacing and element counts for the $Re_\tau=\num{2000}$ channel with mortar meshes.
}
\label{tab:Mortar2}
\centering
\resizebox{!}{0.1075\columnwidth}{\begin{tabular}{|l|ccc|cc|c|c|}
\hline
\multirow{2}{*}{Iteration} &
\multicolumn{3}{c|}{\multirow{2}{*}{$\Delta^+_{i}(y=0)/(P+1)$}} &
\multicolumn{2}{c|}{\texttt{\#Elems}} &
\multirow{2}{*}{\texttt{nDOF}} &
\multirow{2}{*}{$Re_\tau$} \\ \cline{5-6}
&      \multicolumn{3}{c|}{}    & \multicolumn{1}{c|}{total} & saved  &          &         \\ \hline
$l=0$ & 333.33 & 166.67 & 333.33 & \multicolumn{2}{c|}{60}             & 12960    & 569.17  \\
$l=1$ &  277.78 & 46.67  & 166.67 & \multicolumn{2}{c|}{288}            & 62208    & 373.16  \\
$l=2$ &  166.67 & 12.67  & 83.33  & 1320                       & 120    & 285120   & 1841.85 \\
$l=3$ &  104.17 & 4.67   & 38.46  & 7592                       & 1976   & 1639872  & 1723.20 \\
$l=4$ &  59.52  & 2.67   & 16.67  & 27195                      & 18165  & 5874120  & 1875.99 \\
$l=5$ &  26.88  & 2.00   & 9.62   & 142662                     & 128154 & 30814992 & 1967.37 \\ \hline
\end{tabular}}
\end{table}	

The details of the mesh including the resolution and the number of elements saved is listed in \tref{tab:Mortar}. For $N_{i}$ and $\Delta_{i}/h$ we listed the respective values at each iteration for each mortar interface. \tref{tab:Mortar} clearly indicates that using mortars saves us up to approx. 50\% of elements in the simulation compared to a structured mesh with equivalent mortar-free spacing. However, this does not affect the timestep since it is dictated by the smallest cell. For an equivalent load per processor, the overall number of processors of the simulation can however be reduced accordingly. It looks that the higher the iteration, the more we save, since the grid near the wall requires more refinement every time compared to grid in the center of the channel. This case converges to the desired $Re_\tau=\num{2000}$ at $l=5$, if judged only by this metric, with mortars saving almost $50\%$ of the computational cost.

\begin{figure}[!htb]
\centering
\includegraphics{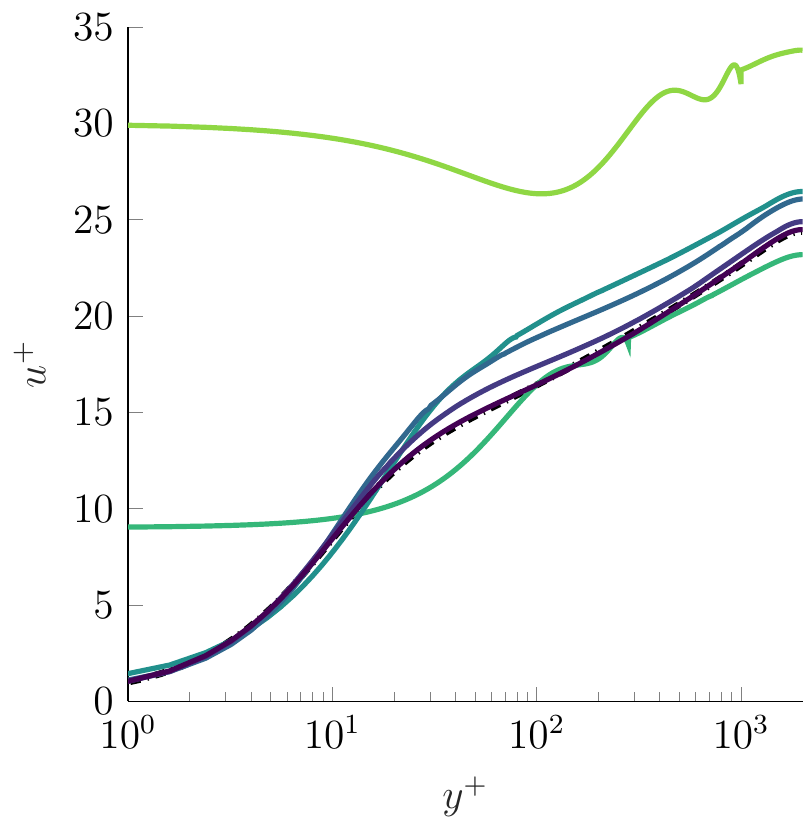}
\includegraphics{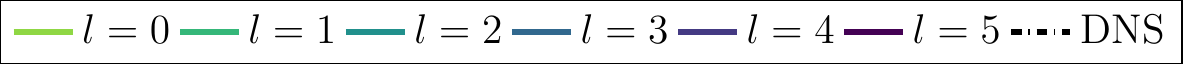}
\caption{Mean velocity profile for the $Re_\tau=\num{2000}$ channel for each iteration $l$.}
\label{fig:umean_Mortar}
\end{figure}

To evaluate the convergence and the quality of the simulations, we again look at the mean velocity and the Reynolds stresses profiles of the channel. The DNS data is found at Lee and Moser \cite{lee:15:channel}, just like the previous case. The mean velocities in \fref{fig:umean_Mortar} show the expected converging behavior. The velocity profile for $l=0$ is  more shifted towards higher friction velocities compared to the $Re_\tau=550$ results in \fref{fig:umean_WRLES}. This can be reasoned with the fact that the $Re_\tau=\num{2000}$ channel is way more underresolved with the same initial mesh $l=0$, and the weak implementation of the boundary conditions therefore cannot generate a good estimation for the $\tau_w$. At iterations $l=[2,3,4,5]$ we see convergence towards the reference data, with $l=5$ matching the DNS velocity profile accurately. 

\begin{figure}[!htb]
\centering
\includegraphics{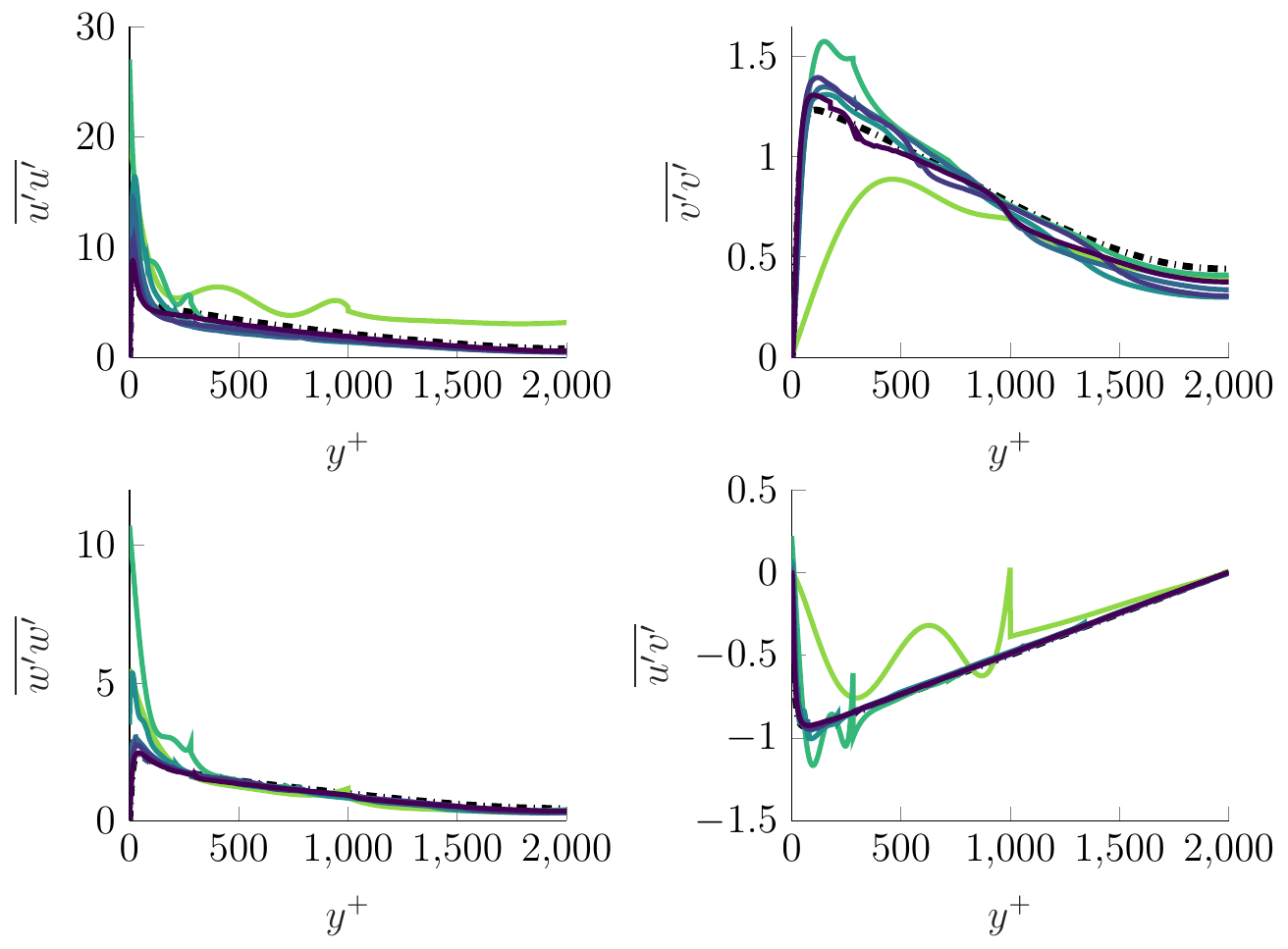}
\includegraphics{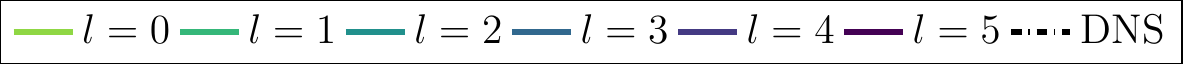}
\caption{Reynolds stresses for the $Re_\tau=\num{2000}$ channel for different iteration $l$.}
\label{fig:reynolds_stresses_Mortar}
\end{figure}

The results for the Reynolds stresses are visualized in \fref{fig:reynolds_stresses_Mortar}. Again, we match the reference solution well for the latest iteration of the grid-adaptation algorithm. The large oscillations for the first iterations are now also more pronounced in $l=1$ in addition to $l=0$ due to the $Re$ being higher compared to the previous case. While $\overline{w'w'}$ and $\overline{u'v'}$ converge well, the results of $\overline{v'v'}$ lack behind. This behavior is expected, since the mortars directly affect the wall-normal directions and thus are visible in the corresponding fluctuations in velocity. Despite that the 	$\overline{v'v'}$ still shows convergence towards the Lee and Moser DNS data, with minimal (but still finite) signs of the mortars at $l=5$. 

\subsection{Channel flow at $Re_\tau = \num{2000}$, application with WMLES}

To conclude the application on channel flow, we run the same test case at $Re_\tau=\num{2000}$ as a wall-modeled simulation. This aims to prove the capabilities of the algorithm to provide a good grid spacing for wall-modeled simulations. We again allow mortar elements in the meshes. Previous experience with the code has shown that a higher $C$ is beneficial in the wall-modeled region, therefore we increase the Vreman constant to $C=0.11$ in Vreman subgrid-scale stress model. We use Spaldings law of the wall as an algebraic wall model, and for its solution we use Newton's method in every Runge-Kutta timestep. The interface height $\hwm$ is fixed at 10\% of the channel half height $h$ and is depicted as a red dashed line in the following figures.

\begin{figure}[!htb]
\centering
\includegraphics{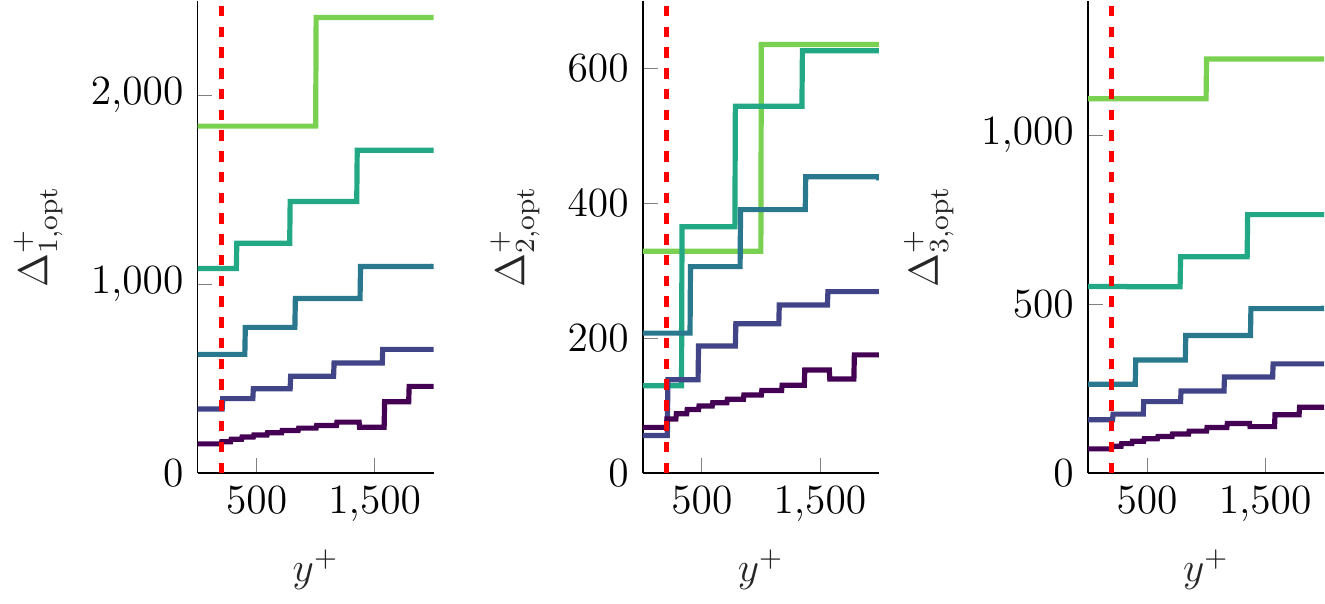}
\includegraphics{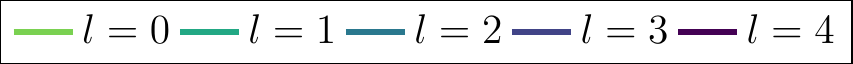}
\caption{Optimal gridspacing according to $G$ for the wall-modeled $Re_\tau=\num{2000}$ channel for different iteration $l$. Dashed red line indicates $\hwm$.}
\label{fig:DeltaX_WMLES}
\end{figure}

First we assess the suggested gridspacings $\Delta^+_{\text{i,{\rm opt}}}$, shown in \fref{fig:DeltaX_WMLES}. The spacing in wall-normal direction is again realized by the same method as for the previous subsection of WRLES containing mortars. However, we have adapted the suggested grid spacings to the WMLES methodology. The pure grid-adaptation finds that the underresolution of the relevant scales is harshest near the wall, thus suggesting the most refinement there. However, the point of wall-modeling is to model the smaller but flow-relevant eddies below $\hwm$, so the grid-adaptation process needs to be told that severe underresolution in that region is a choice and not an error. To achieve this, we introduce a restriction on $\Delta^+_{\text{i,{\rm opt}}}$ according to Toosi (personal communication). 
We do not allow any smaller $\Delta^+_{\text{i,{\rm opt}}}$ below $\hwm$, meaning that we reset the values of $\Delta^+_{\text{i,{\rm opt}}}$ for the cells that lie between $\hwm$ and the wall to the values of their counterparts at $\hwm$. This applies to the grid spacings of all directions in space. 
\fref{fig:DeltaX_WMLES} also indicates that the wall model interface is not restricted to the wall neighboring grid cell, but can also be placed in the second cell, \eg $l=4$. For $l=4$ we also introduce a mortar in $x$ and $z$ which is located at the same wall-normal positions for both dimensions. In comparison to the wall-resolved case this only happens at the fourth iteration, since the scale differences between wall and the center of the channel are much smaller, since wall-modeled LES does not resolve the small scales of the wall turbulence. 
The resolution and resulting viscous spacings at the wall are listed in \tref{tab:WMLES}. 

\begin{table}[]
\caption{Simulation properties of the wall-modeled $Re_\tau=550$ channel for different iteration $l$.
}
\label{tab:WMLES}
\centering
\resizebox{\columnwidth}{!}{\begin{tabular}{|l|ccccc|ccccc|ccc|cc|c|c|}
\hline
\multirow{2}{*}{Iteration} &
\multicolumn{2}{c}{\multirow{2}{*}{$N_1$}} &
\multicolumn{1}{c}{\multirow{2}{*}{$N_2$}} &
\multicolumn{2}{c|}{\multirow{2}{*}{$N_3$}} &				
\multicolumn{2}{c}{\multirow{2}{*}{$\Delta_\text{1}/h$}} &
\multicolumn{1}{c}{\multirow{2}{*}{$(\Delta_\text{2}/h)_\text{min}$}} &
\multicolumn{2}{c|}{\multirow{2}{*}{$\Delta_\text{3}/h$}} &
\multicolumn{3}{c|}{\multirow{2}{*}{$\Delta^+_{\text{i}}(y=0)/(P+1)$}} &
\multicolumn{2}{c|}{\texttt{\#Elems}} &
\multirow{2}{*}{\texttt{\#DOF}} &
\multirow{2}{*}{$Re_\tau$} \\ \cline{15-16}
&
\multicolumn{5}{c|}{} &
\multicolumn{5}{c|}{} &
&
\multicolumn{2}{c|}{} &
\multicolumn{1}{c|}{total} &
saved &
&
\\ \hline
$l=0$ &
\multicolumn{2}{c}{10} &
\multicolumn{1}{|c|}{2} &
\multicolumn{2}{c|}{3} &
\multicolumn{2}{c}{1.00} &
\multicolumn{1}{|c|}{0.50} &
\multicolumn{2}{c|}{1.00} &
333.33 &
166.67 &
333.33 &
\multicolumn{2}{c|}{60} &
12960 &
627.91 \\
$l=1$ &
\multicolumn{2}{c}{12} &
\multicolumn{1}{|c|}{4} &
\multicolumn{2}{c|}{5} &
\multicolumn{2}{c}{0.83} &
\multicolumn{1}{|c|}{0.25} &
\multicolumn{2}{c|}{0.60} &
277.78 &
54.67 &
200.00 &
\multicolumn{2}{c|}{240} &
51840 &
872.29 \\
$l=2$ &
\multicolumn{2}{c}{19} &
\multicolumn{1}{|c|}{5} &
\multicolumn{2}{c|}{11} &
\multicolumn{2}{c}{0.53} &
\multicolumn{1}{|c|}{0.20} &
\multicolumn{2}{c|}{0.27} &
175.44 &
21.33 &
90.91 &
\multicolumn{2}{c|}{1045} &
225720 &
1241.07 \\
$l=3$ &
\multicolumn{2}{c}{32} &
\multicolumn{1}{|c|}{6} &
\multicolumn{2}{c|}{23} &
\multicolumn{2}{c}{0.31} &
\multicolumn{1}{|c|}{0.17} &
\multicolumn{2}{c|}{0.13} &
104.17 &
34.67 &
43.48 &
\multicolumn{2}{c|}{4416} &
953856 &
1220.01 \\
$l=4$ &
62 &
31 &
\multicolumn{1}{|c|}{14} &
38 &
19 &
0.16 &
0.32 &
\multicolumn{1}{|c|}{0.07} &
0.08 &
0.16 &
53.76 &
10.00 &
26.32 &
31217 &
1767 &
6742872 &
1555.60 \\ \hline
\end{tabular}}
\end{table}

\begin{figure}[!htb]
\centering
\includegraphics{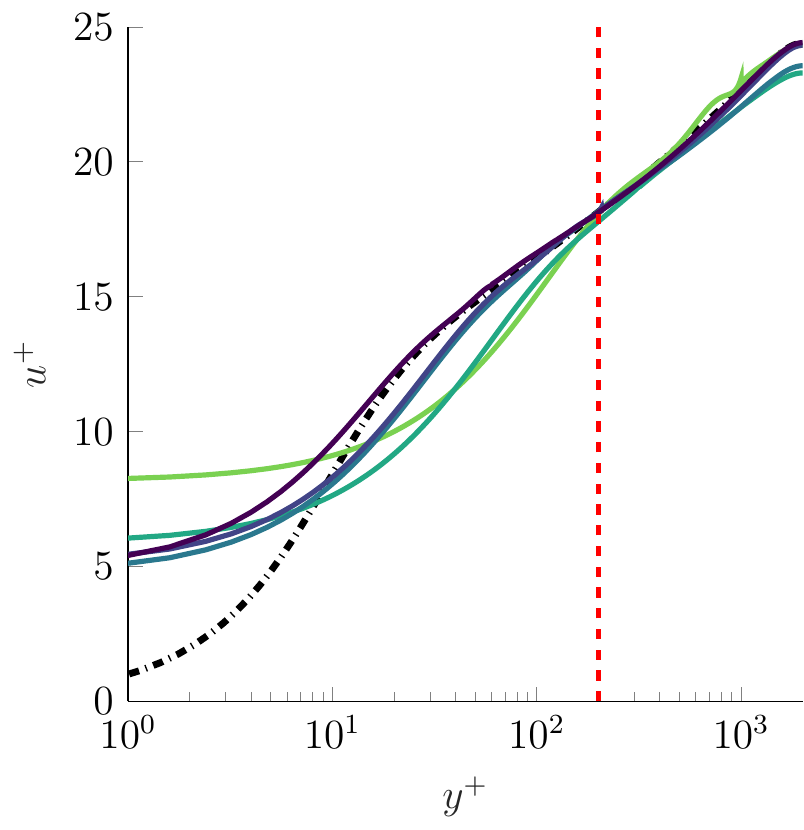}
\includegraphics{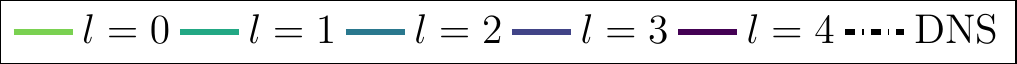}
\caption{Mean velocity profile for the wall-modeled $Re_\tau=\num{2000}$ channel for each iteration $l$.}
\label{fig:umean_WMLES}
\end{figure}

To verify the results, we again compare them against the same DNS data by Lee and Moser. In \fref{fig:umean_WMLES} even for $l=0$ we already have good agreement, especially at the interface location at $h^+_{wm}=200$. This is because the initial grid is much closer to the WMLES requirements than the WRLES requirements. A clear convergence towards the DNS data is also visible. Iteration $l=3$ agrees with the DNS data for the whole channel above $\hwm$, including the center of the channel which was slightly off for the previous iterations.

\begin{figure}[!htb]
\centering
\includegraphics{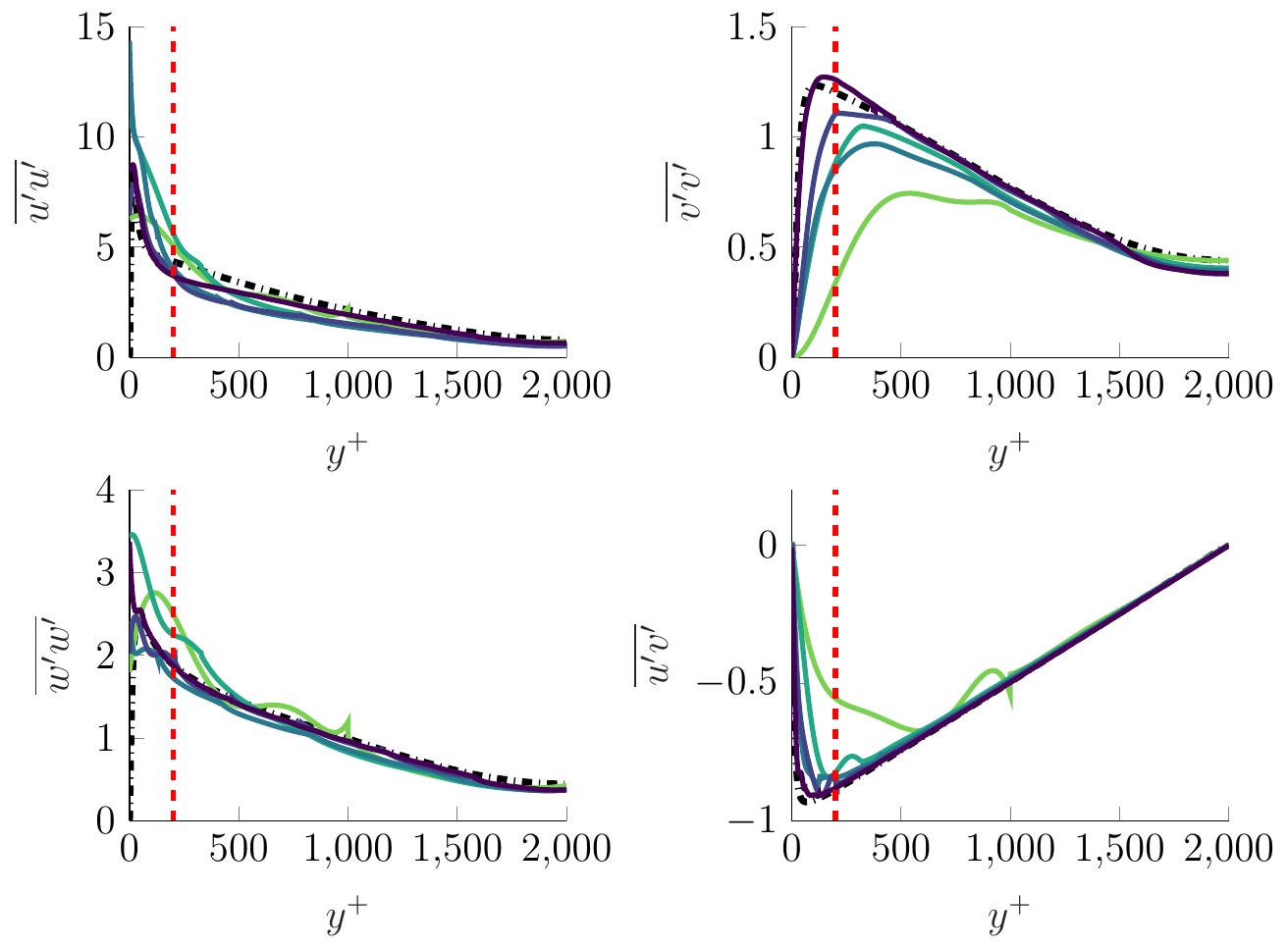}
\includegraphics{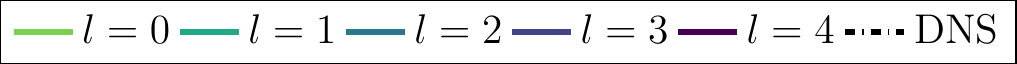}
\caption{Reynolds stresses for the wall-modeled $Re_\tau=\num{2000}$ channel for different iteration $l$.}
\label{fig:reynolds_stresses_WMLES}
\end{figure}

The Reynolds stresses are depicted in \fref{fig:reynolds_stresses_WMLES}. Unlike the velocity profiles, the fourth iteration is needed to match the Reynolds stresses better. This is due to the WMLES approach, not the grid-adaptation. Even then, the Reynolds stresses, especially $\overline{u'u'}$ does not exactly converge on the DNS data, but become very acceptable for a WMLES. The $\overline{v'v'}$ also deserves some attention, where we can clearly see the converging behavior of the fluctuations; but in the center of the channel the mortar in the mesh of $l=4$ gets visible again. The other Reynolds stress components show good enough agreement, proving the applicability of the algorithm to WMLES.

\section{Application to the flow over an airfoil}

To conclude this work, we investigate the flow over a transonic NACA 64A-110 airfoil at an angle of attack of $\alpha=0\deg$. This test case allows us to apply all the functionalities that we introduced in the channel flow onto a more practical application. The airfoil grid is generated by extruding a 2D geometry ($xy$-plane) in $z$. The $xy$-plane grid is generated using an unstructured 2D mesh tool. Due to the extrusion the mesh will be structured in $z$. To allow the grid-adaptation algorithm to show its full potential, we allow a homogeneous mortar element distribution along $z$ by specifying regions in the $xy$-plane with different number of elements in $z$. This way, we generate a fully unstructured, yet still hexahedral only, three dimensional grid. The characteristic length of the airfoil simulation $c$ is set to the chord length and defined as $c=1$. The spanwise extension in $z$ is $L_z=0.05c$.

The airfoil is placed in a wind tunnel, therefore we include Euler walls above and below the profile. The dimensionless simulation parameters are $Ma=0.72$ and $Re_c=\num{930000}$. We perform it as a WMLES. The reliable transition to turbulence is achieved through the flow being tripped with a numerical trip at $x=0.05c$ using the method introduces by Schlatter and Örlu \cite{schlatter2012} on the suction and pressure side. The interface height is fixed at $\hwm=0.0134$ and constant over the surface for all iterations and has been determined with the interface height adaptation algorithm introduced by Kahraman et al. \cite{kahraman2022} on the WRLES reference data. 

We eventually compare the results to a wall-resolved LES simulation that has been performed previously. We start with a coarse mesh and perform the iterative approach, just like the channel test case. We iterate until we match the pressure coefficient $c_p$, the skin friction $\tau_w$ and the boundary layer thickness $\delta_{99}$.

\begin{figure}[!htb]
\centering
\begin{subfigure}[b]{0.49\textwidth}
\centering
\includegraphics[width=0.9\columnwidth]{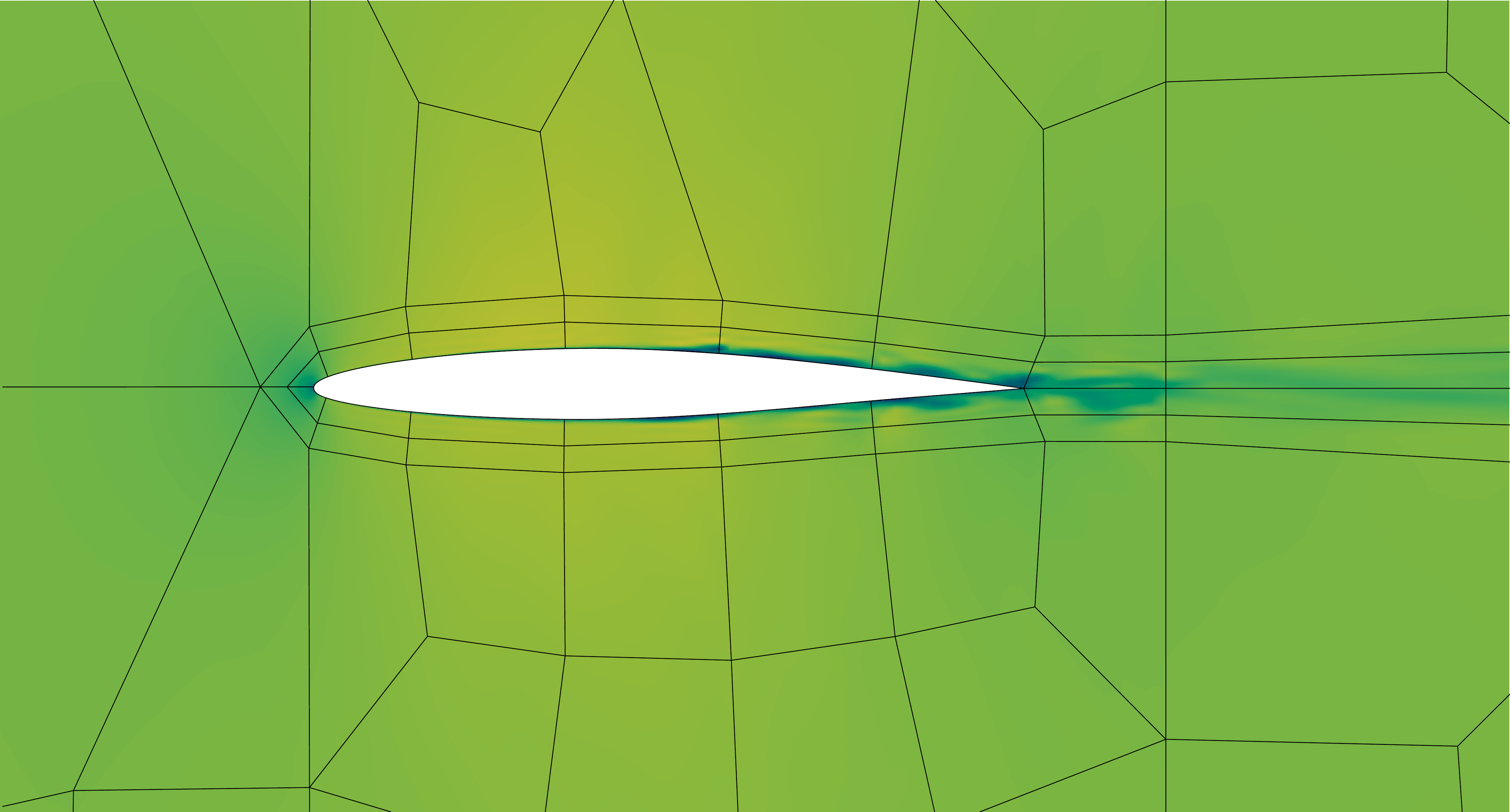}
\caption{initial mesh $l=0$ around the airfoil}
\label{fig:level_0}
\end{subfigure}
\hfill
\begin{subfigure}[b]{0.49\textwidth}
\centering
\includegraphics[width=0.9\columnwidth]{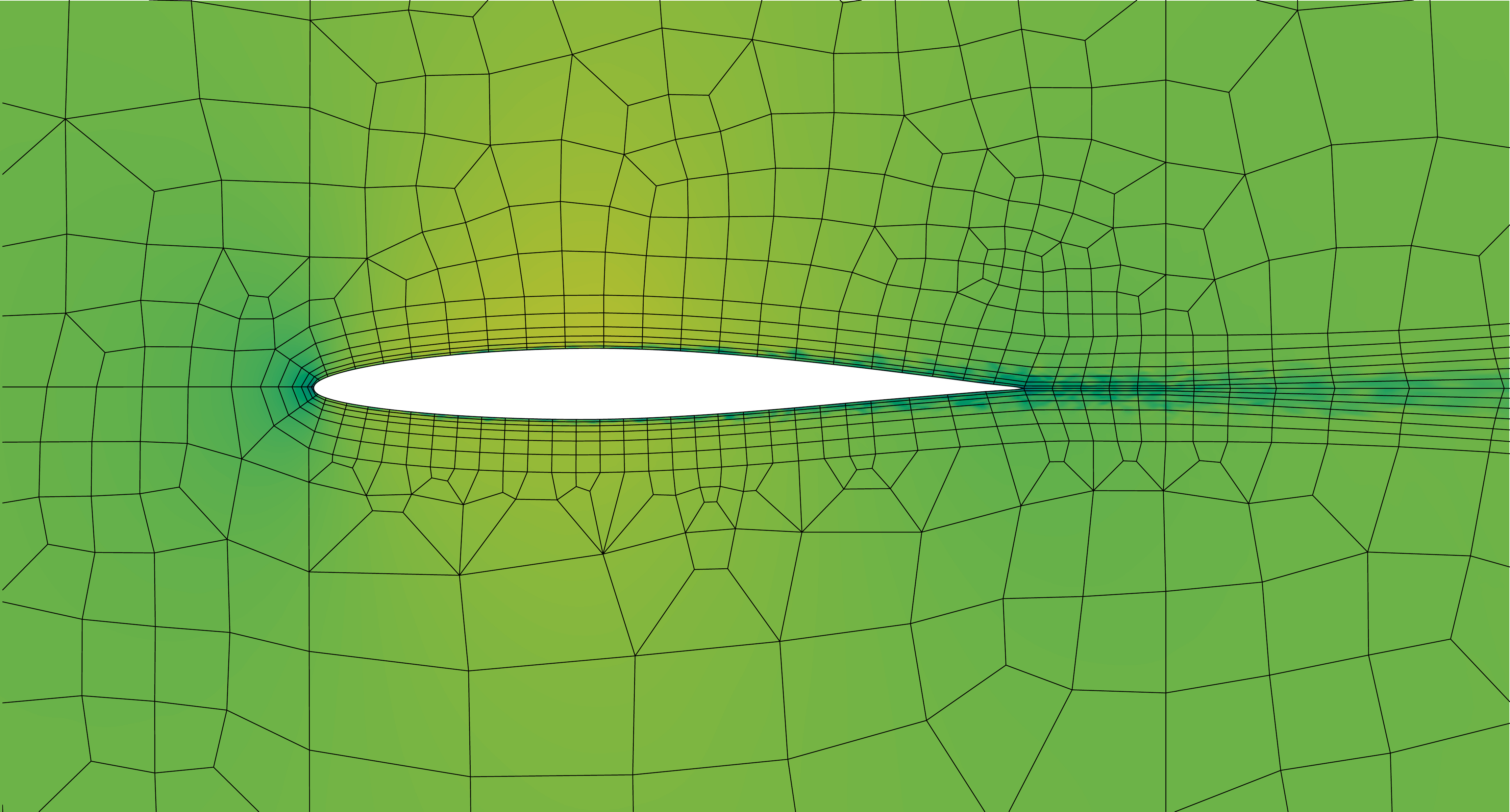}
\caption{first refinement level $l=1$}
\label{fig:level_1}
\end{subfigure}
\begin{subfigure}[b]{0.98\textwidth}
\centering
\includegraphics[width=0.3\columnwidth]{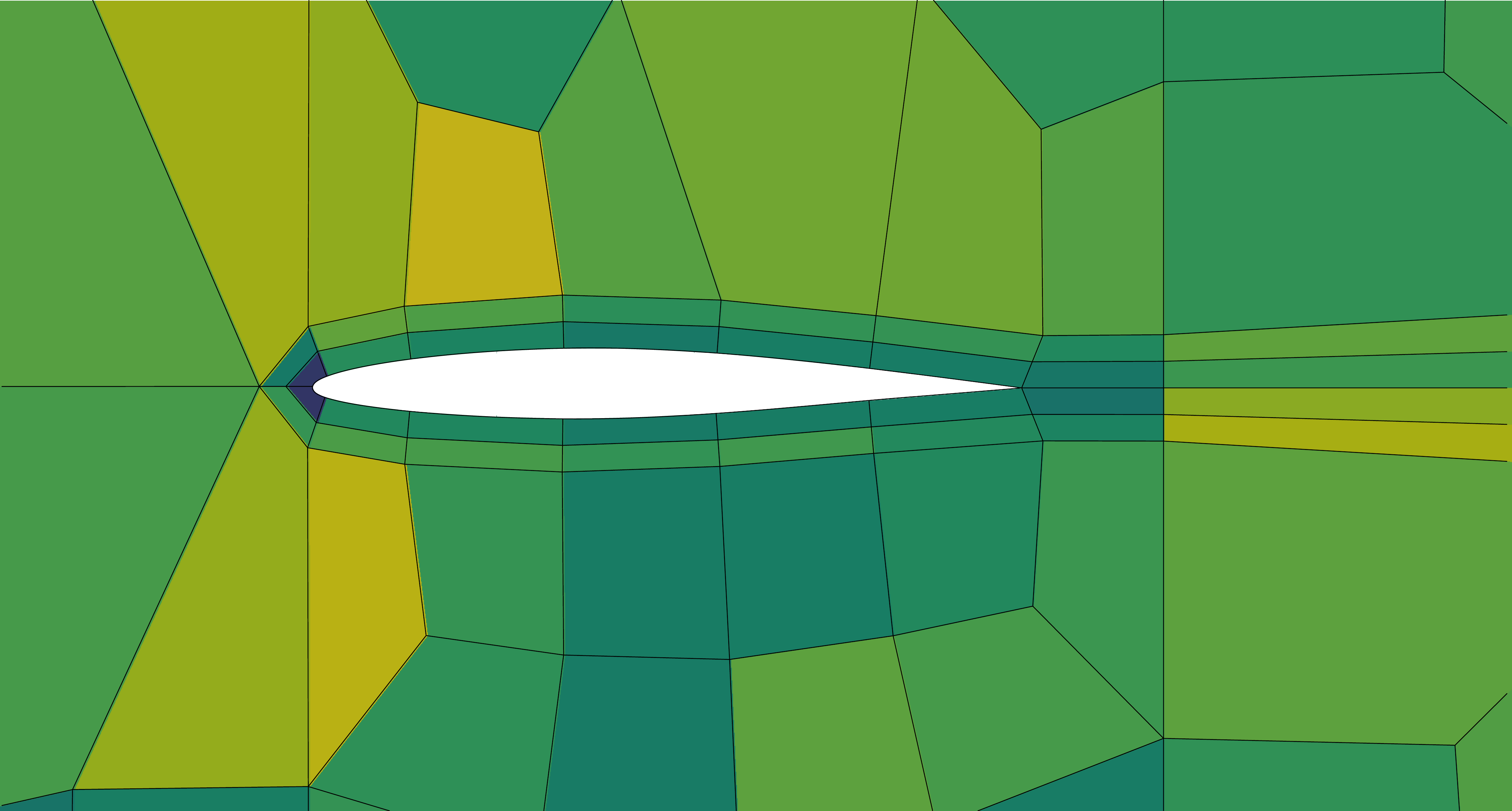}
\hfill
\includegraphics[width=0.3\columnwidth]{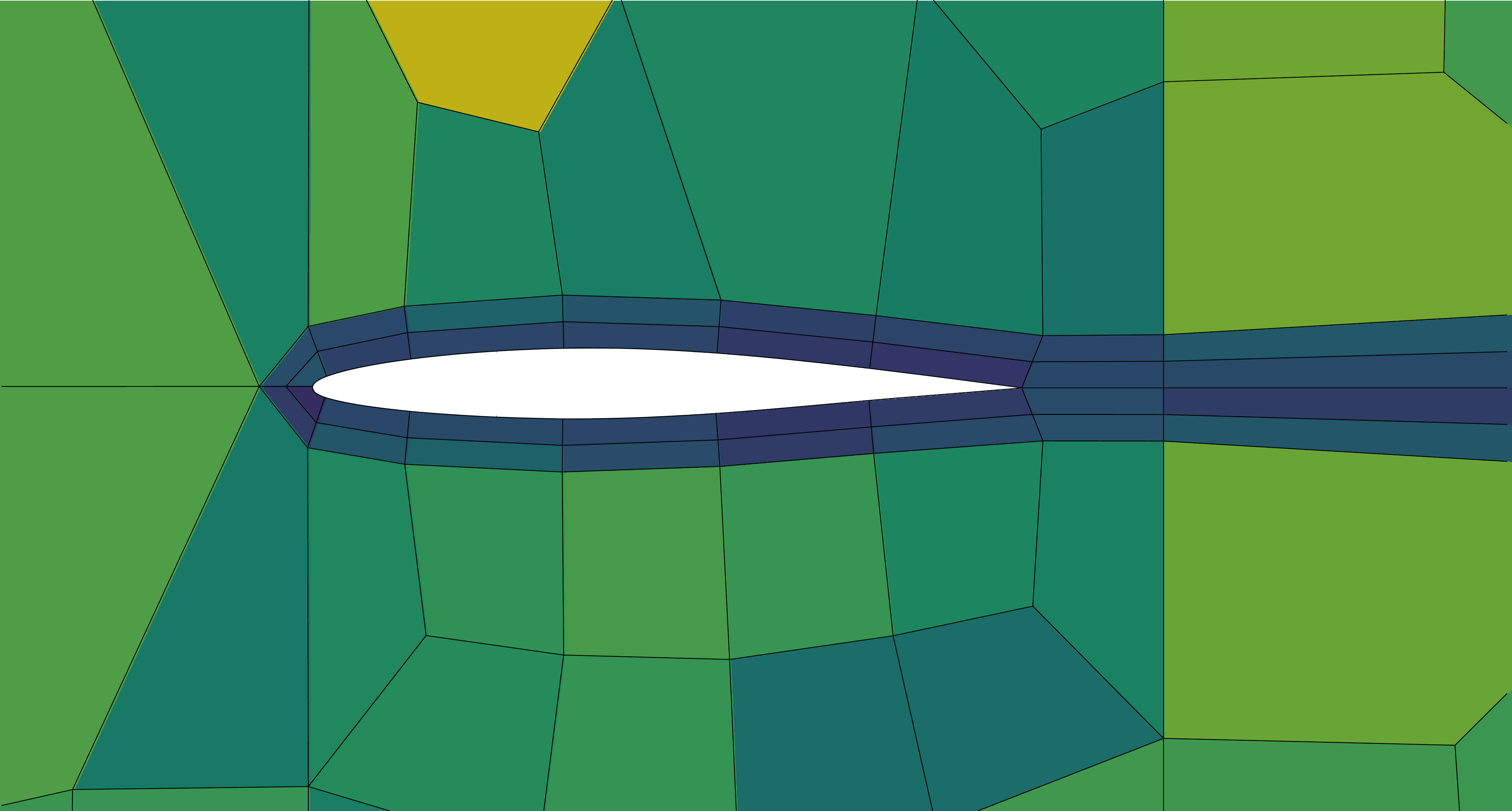}
\hfill
\includegraphics[width=0.3\columnwidth]{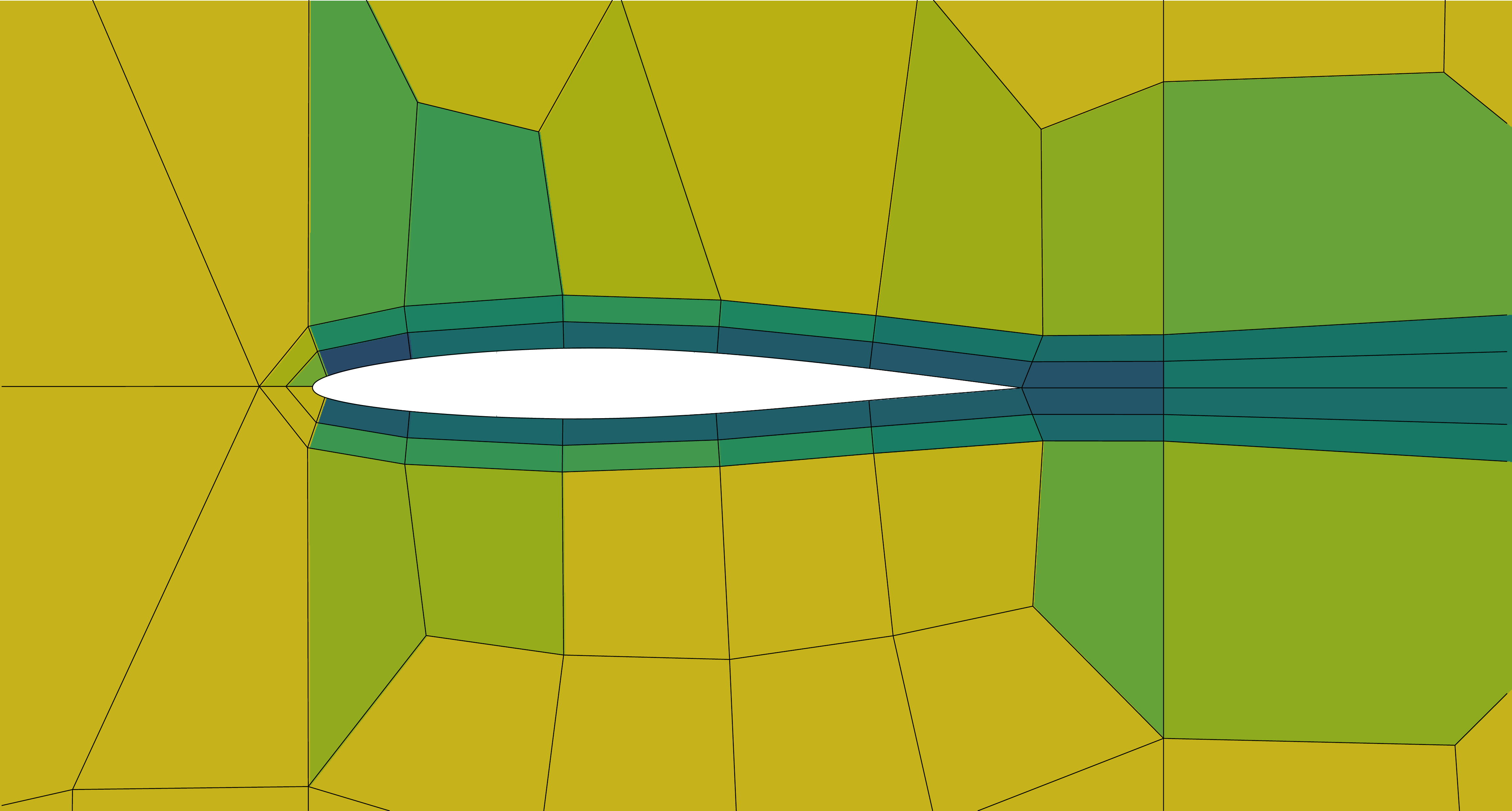}
\hfill
\resizebox{0.07\columnwidth}{!}{
\begin{picture}(165,384.907347)\Huge
	\put(0,369.246474){\num{5.0E-1}}\put(0,319.961315){\num{2.0E-1}}\put(0,276.47819){\num{1.0E-1}}\put(0,234.355951){\num{5.0E-2}}\put(0,177.774588){\num{2.0E-2}}\put(0,135.036079){\num{1.0E-2}}\put(0,091.852824){\num{5.0E-3}}\put(0,034.634344){\num{2.0E-3}}\put(0,0){\num{1.0E-3}}\put(65,0){\includegraphics[width=100pt,page=1]{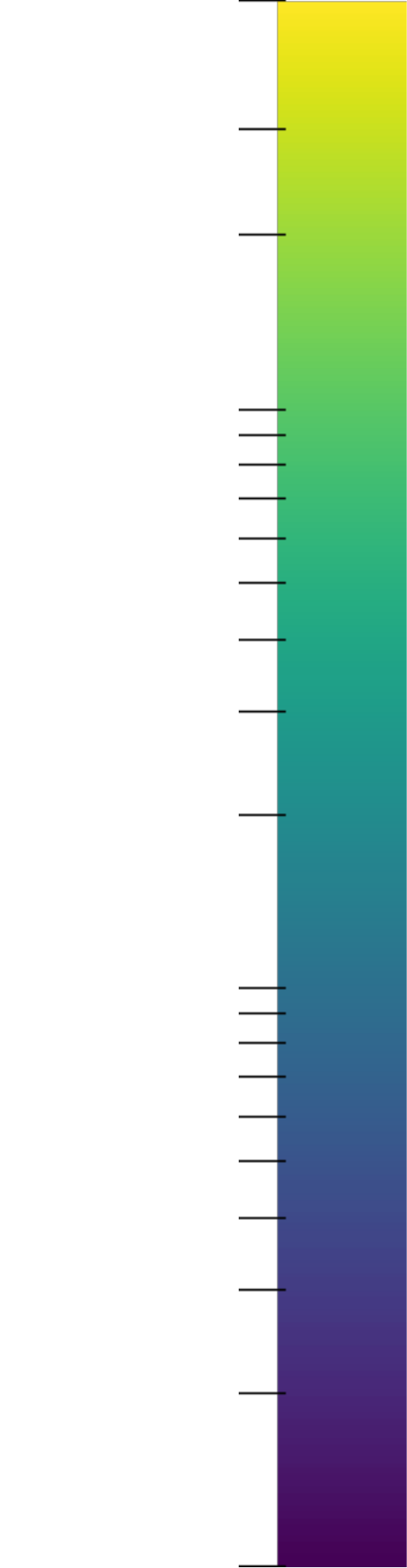}}\end{picture}}
\caption{Qualitative contour plot of $\Delta_\text{i,opt}$ with logarithmic scaling after $l=0$. All plots are scaled equivalently. From left to right: wall parallel $s$, wall normal $n$ and spanwise direction $z$.}
\label{fig:DeltaXOpt_airfoil}
\end{subfigure}

\caption{Comparison between the velocity field and mesh the initial and the first iteration.}
\label{fig:image_airfoil}
\end{figure}
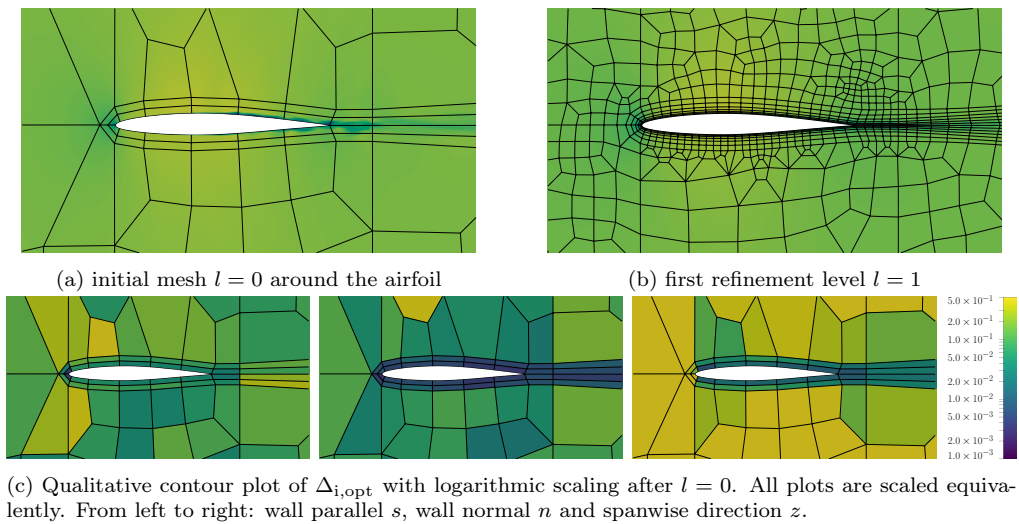

The initial mesh $l=0$ is built to be extremely cheap and without any special treatment of the boundary layer or any other flow properties. The only constraint on the mesh is to fit the geometry. Thus in $l=0$ the mesh has only 6 elements per cordlength in streamwise direction, which results in a maximum $y^+\approx1100$. This initial mesh is depicted in \fref{fig:level_0}. Since we use high-order methods we must use curved elements, which allows to take more details of the geometry into account even for as coarse meshes as in $l=0$.

To get the next mesh, of iteration $l=1$, we evaluate the grid error indicator densities $g(\vec{x},\vec{n}_i)$ to get $\Delta_{i,{\rm opt}}$, which is qualitatively visualized in \fref{fig:DeltaXOpt_airfoil}. It shows the same behavior that was already observed for the channel, that we refine the most in wall normal direction $n$,
followed by the spanwise direction $z$ and the tangential direction $s$. Interesting to note is that the spanwise direction has very coarse suggested grid spacing at the leading edge. This is expected, and nice to see for the performance of the algorithm, since that the flow is laminar which means the lengthscale approaches infinity along that direction. The turbulent wake also is refined in more detail, since this coarse grid is still able to sustain some turbulence there. The mesh $l=1$ is also visualized in \fref{fig:level_1}, showing significantly more elements near the airfoil, which is very well expected. Fig \ref{fig:image_airfoil} thus shows the complete process from $l=0$ to $l=1$. 

\begin{figure}[!htb]
\centering
\includegraphics{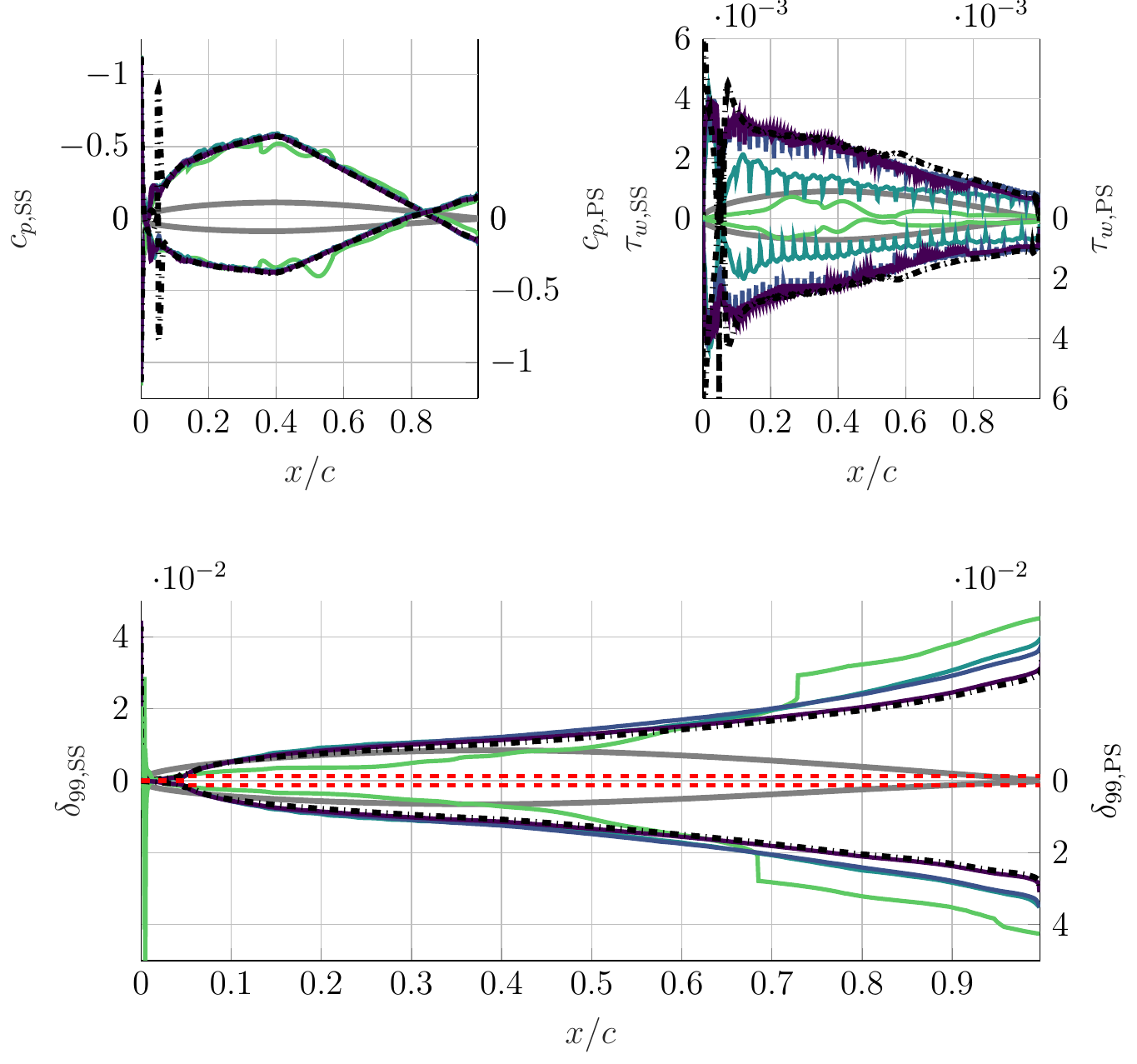}
\includegraphics{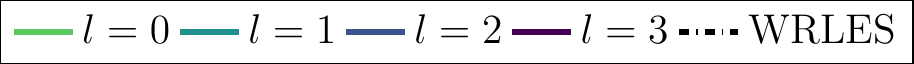}	
\caption{Convergence of pressure coefficient $c_p$ and skin friction $\tau_w$ and boundary layer thickness $\delta_{99}$ on the pressure (PS) and suction (SS) side of the NACA 64A-110 airfoil.}
\label{fig:airfoil_convergence}
\end{figure}

The whole development of the relevant quantities defined earlier is visualized in \fref{fig:airfoil_convergence}. The graphs indicate a clearly converging behavior and also shows the different rates of convergence of the quantities. The pressure coefficient $c_p$ converges the fastest. The skin friction coefficient $c_f$ only converges after $l=2$ and the boundary layer thickness, which is most sensitive to the grid resolution, needs four iterations to converge. The difference between $l=2$ and $l=3$ is only the resolution in $x$ and $z$, which is listed in \tref{tab:airfoil_1} and \tref{tab:airfoil_2} along other grid characteristics. Especially at the trailing edge the algorithm suggested mesh refinement. This is also where the discrepancy between WRLES reference data and the iterations $l=[0,1,2]$ was the largest. The red dashed line in \fref{fig:airfoil_convergence} shows the wall model interface height over $x/c$ which is roughly 10\% of $\delta_{99,\text{mean}}$.

\begin{table}[]
\caption{Grid spacings of the wall-modeled NACA 64A-110 airfoil for different iteration $l$.
}
\label{tab:airfoil_1}
\centering
\resizebox{\columnwidth}{!}{\begin{tabular}{|l|ccc|llllll|}
\hline
\multirow{2}{*}{Iteration} &
\multicolumn{3}{c|}{\multirow{2}{*}{$\Delta^+_\text{SS}(x=0.05c,s)/(P+1)$}} &
\multicolumn{6}{c|}{$\Delta_\text{i}/c$} \\ \cline{5-10}
\multicolumn{1}{|c|}{} &
\multicolumn{3}{c|}{} &				
\multicolumn{1}{c}{min.} &
\multicolumn{1}{c|}{max.} &
\multicolumn{1}{c}{min.} &
\multicolumn{1}{c|}{max.} &
\multicolumn{1}{c}{min.} &
\multicolumn{1}{c|}{max.} \\ \hline
$l=0$ &
1267.21 &
219.82 &
146.54 &
1.21e-1 &
\multicolumn{1}{l|}{2.16e-1} &
3.75e-2 &
\multicolumn{1}{l|}{3.75e-2} &
\multicolumn{2}{c|}{2.50e-2}  \\
$l=1$ &
316.92 &
52.76 &
48.85 &
3.30e-2 &
\multicolumn{1}{l|}{5.40e-2} &
9.00e-3 &
\multicolumn{1}{l|}{9.00e-3} &
8.33e-3 &
1.67e-2 \\
$l=2$ &
146.54 &
5.86 &
12.21 &
2.10e-2 &
\multicolumn{1}{l|}{3.00e-2} &
1.00e-3 &
\multicolumn{1}{l|}{2.00e-3} &
2.08e-3 &
1.67e-2\\
$l=3$ &
58.62 &
5.86 &
12.21 &
5.00e-3 &
\multicolumn{1}{l|}{1.50e-2} &
1.00e-3 &
\multicolumn{1}{l|}{2.50e-3} &
1.04e-3 &
1.67e-2\\ \hline
\end{tabular}}
\end{table}

\begin{table}[]
\caption{Number of cells of the wall-modeled NACA 64A-110 airfoil for different iteration $l$.}
\label{tab:airfoil_2}
\centering
\resizebox{0.7\columnwidth}{!}{\begin{tabular}{|l|cc|cc|c|}
\hline
\multirow{2}{*}{Iteration} &
\multicolumn{2}{c|}{$N_i$} &
\multicolumn{2}{c|}{\texttt{\#Elems}} &
\multirow{2}{*}{\texttt{\#DOF}} \\ \cline{2-5}
&
\multicolumn{1}{c|}{$N_s\cdot N_n$} &
\multicolumn{1}{c|}{$N_z$} &
\multicolumn{1}{c|}{total} &
saved &
\\ \hline
$l=0$ &
395 &
$(2)$ &
790 &
&
404480 \\
$l=1$ &
1472 &
$(3,6)$ &
5946 &
2886 &
3044352 \\
$l=2$ &
2590 &
$(3,6,12,24)$ &
36072 &
26088 &
18468864 \\
$l=3$ &
9111 &
$(3,6,12,24,48)$ &
222426 &
214902 &
113882112 \\ \hline
\end{tabular}}
\end{table}

Table \ref{tab:airfoil_2} also shows the mortar information for $l=[1,2,3]$ as well as the minimal and maximum grid spacings at the airfoil boundaries. The values in between can be easily calculated by assuming the 2-1 mortar interfaces. The observation made in \tref{tab:Mortar} are confirmed here. The usage of the unstructured capabilities of the discontinuous Galerkin framework allow us to save up to approximately 50\% of the elements, which directly translates to the same amount of savings in compute time. 

\begin{figure}[!htb]
\centering
\begin{subfigure}[b]{0.49\textwidth}
\centering
\includegraphics[width=0.9\columnwidth]{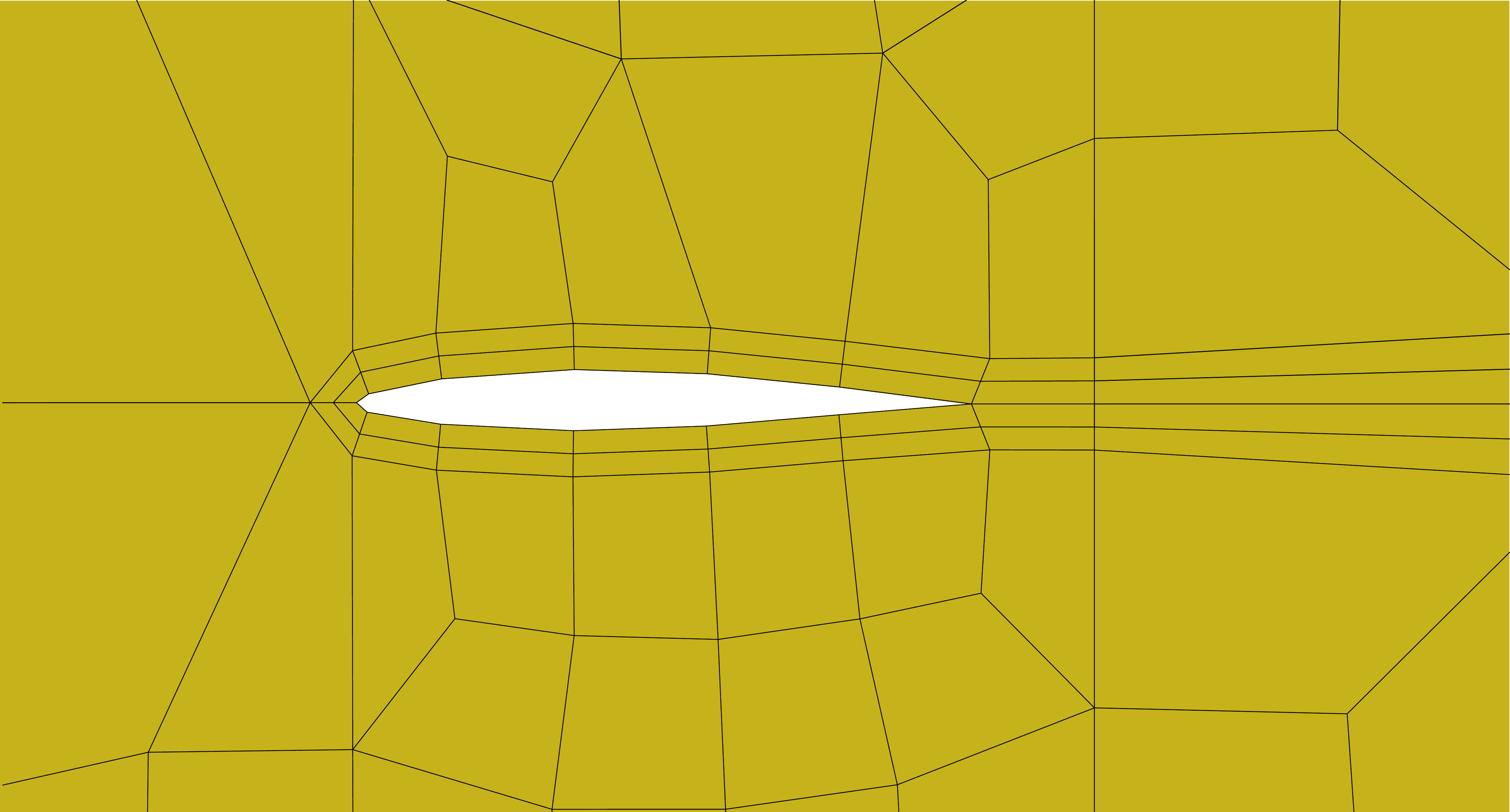}
\caption{Initial mesh $l=0$ around the airfoil.}
\label{fig:zone_0}
\end{subfigure}
\hfill
\begin{subfigure}[b]{0.49\textwidth}
\centering
\includegraphics[width=0.9\columnwidth]{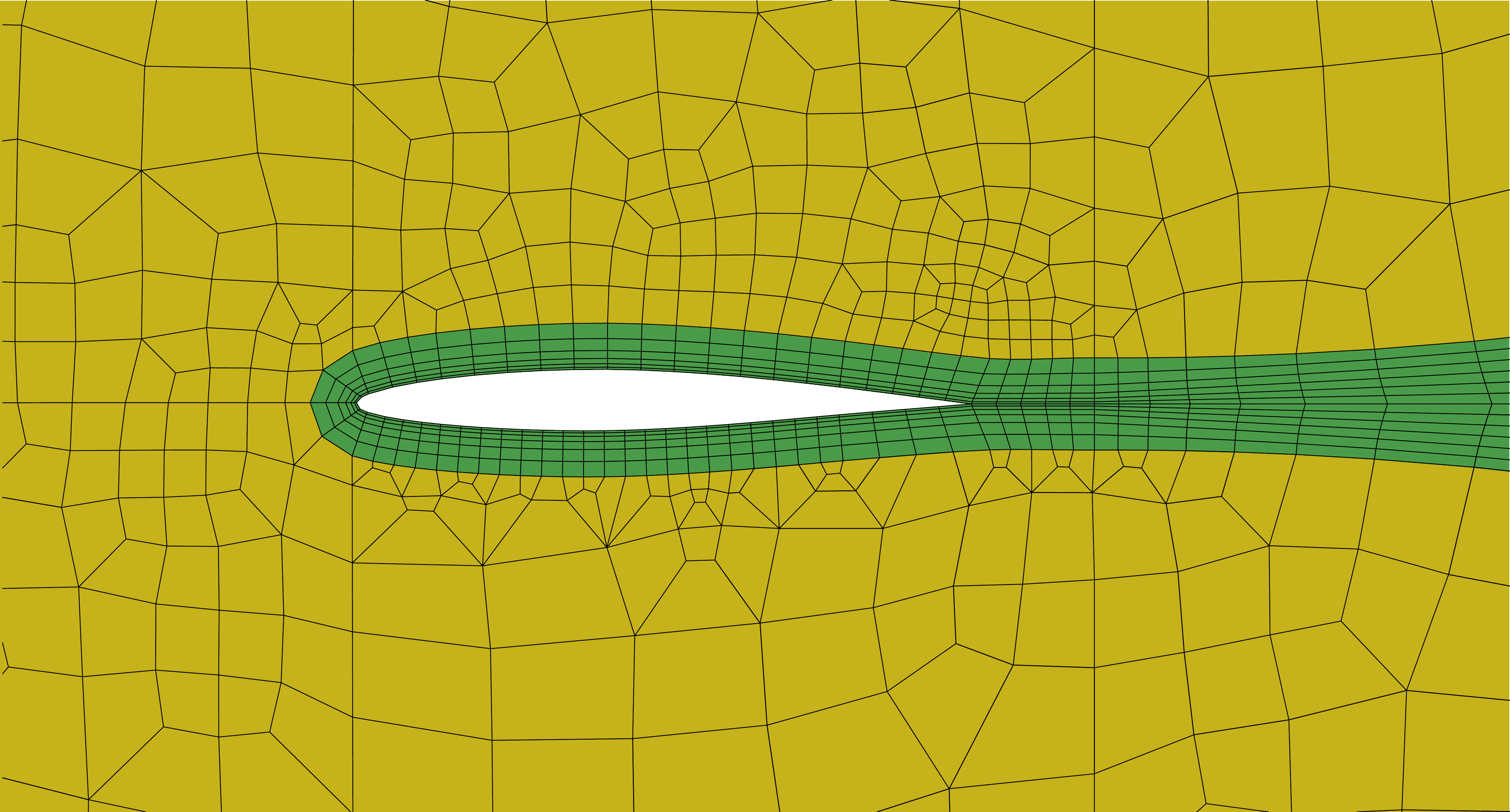}
\caption{First refinement iteration $l=1$.}
\label{fig:zone_1}
\end{subfigure}
\vskip\baselineskip
\begin{subfigure}[b]{0.49\textwidth}
\centering
\includegraphics[width=0.9\columnwidth]{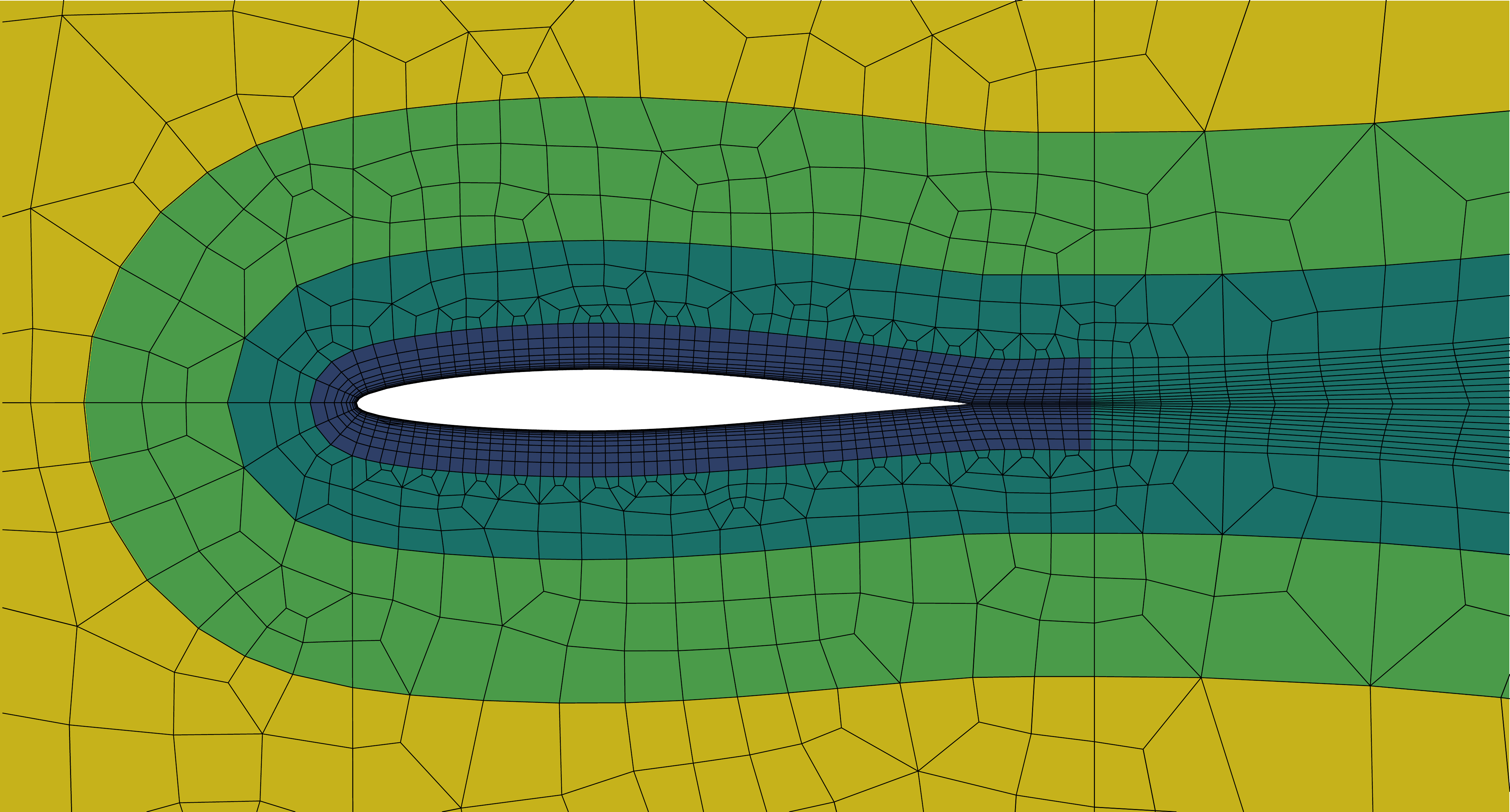}
\caption{Second refinement iteration $l=2$.}
\label{fig:zone_2}
\end{subfigure}
\hfill
\begin{subfigure}[b]{0.49\textwidth}
\centering
\includegraphics[width=0.9\columnwidth]{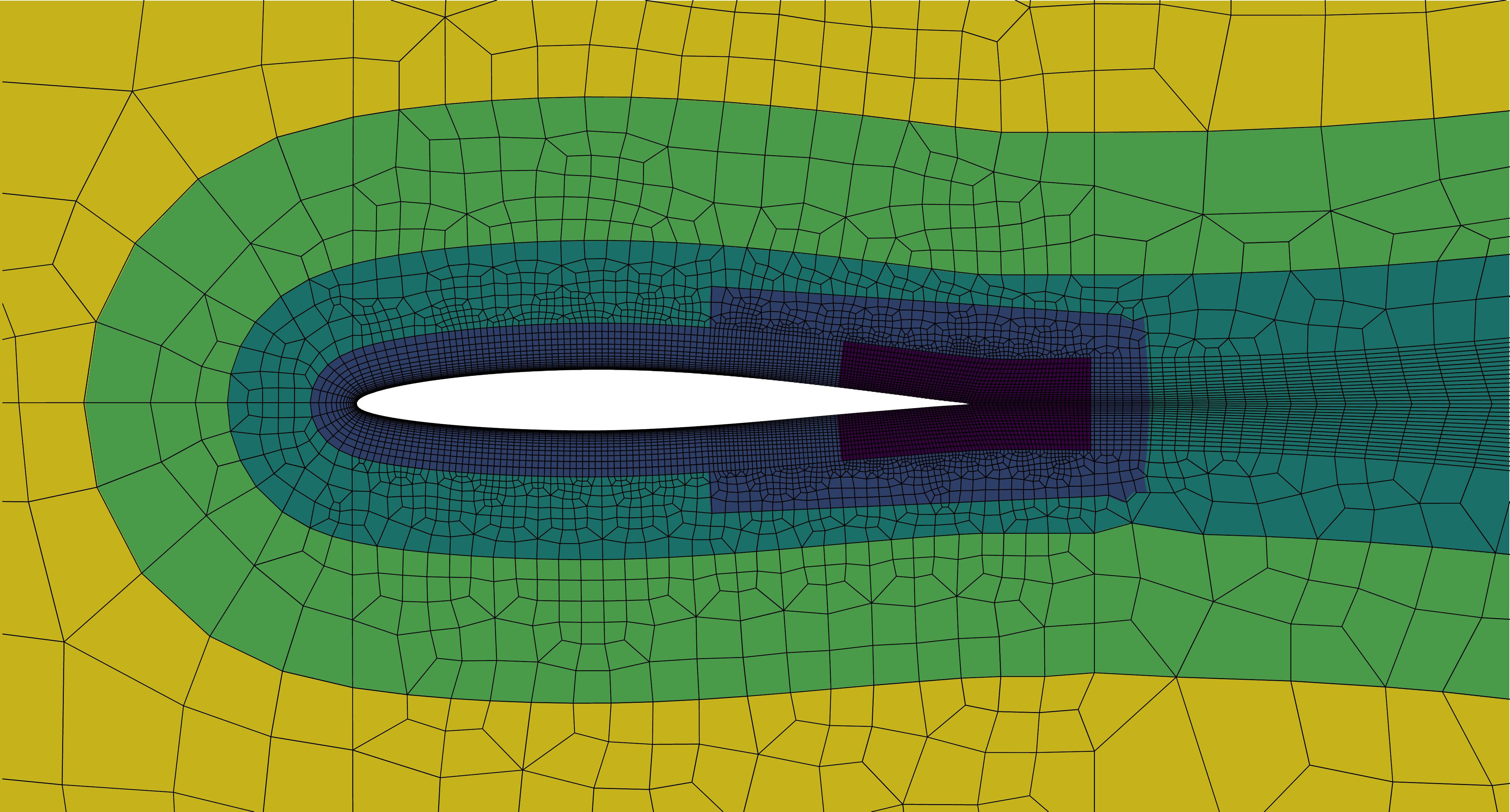}
\caption{Third refinement iteration $l=3$.}
\label{fig:zone_3}
\end{subfigure}
\includegraphics{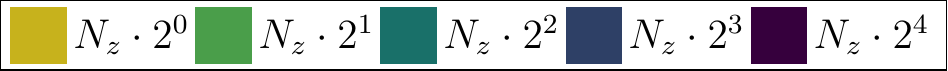}
\caption{Comparison between the $z$ mortar zones of the airfoil iterations, visualized with non-curved elements.}
\label{fig:zones_airfoil}
\end{figure}

In \fref{fig:zones_airfoil} the mortar zones in $z$ are visualized, each denoting a different number of elements in $z$. The colors are based on the minimum number of cells along spanwise direction $N_z$, which are $N_z(l=0) = 2$ and $N_z(l=[1,2,3]) = 3$. The zones are designed in a way to meet the target values of the indicator in an efficient way with our limitation to 2-1 mortar interfaces, even though the indicator suggests even more coarsening in some locations and much more rapid refinement at other locations, \eg from $N_z\cdot 2^4$ to $N_z\cdot 2^0$. We insert buffer layers in between to make the mesh compatible and to enable a transition between the mortar zones. Still, the number of elements saved justifies this approach.

\section{Summary}

In this work we presented turbulent channels and a transonic airfoil in discontinuous Galerkin framework using different grids proposed by using the high-order optimized version of the grid-adaptation algorithm based on the works from Toosi and Larsson \cite{toosi:20, toosi:21}. The boundary layers were either wall-resolved or wall-modeled to show the ability of the algorithm to handle two major cases of large eddy simulation. For all test cases we showed convergence and good agreement with the results in literature.

The grid-adaptation indicator in general for high-order methods has turned out to be most efficient when applied elementwise and not per collocation point due to the reduced robustness of the numerical scheme in underresolved settings. Having the indicator values on the element, splitting and remeshing with the proposed recommended gridspacing is straight forward. The application of the grid-adaptation machinery to a $p$-refined approach is yet to be investigated.

Another important aspect of the simulation run is the usage of the unstructured capabilities of the discontinuous Galerkin implementation in the FLEXI framework. Unstructured meshes in three dimensions were realized using mortar interfaces between elements and the usage of these elements saved up to 50\% of compute time without any significant loss in accuracy. Especially for the airfoil case, we were able to clearly distinguish different mortar zones in spanwise direction, even after few iterations. Therefore, we have shown that the grid-adaptation framework is able to deliver a problem-tailored and cost-optimized mesh with converging behavior.

Thus, the grid-adaptation algorithms can be applied to arbitrary geometries without any prior knowledge to the flow field and still create a problem-tailored and cost optimized mesh, which does not rely on the experience of the user.

\section*{Acknowledgments}

MB and AB gratefully acknowledge the Deutsche Forschungsgemeinschaft DFG (German Research Foundation) for funding this work in the framework of the research unit FOR2895, and thank the Gauss Centre for Supercomputing e.V. (www.gauss-centre.eu) for funding this project (GCS-lesdg) by providing computing time on the GCS Supercomputer HAWK at Höchstleistungsrechenzentrum Stuttgart (www.hlrs.de).
AK and JL were supported by the Department of Energy PSAAP III program (grant DE-NA0003993) and the NASA Transformational Tools and Technologies project (grant 80NSSC18M0148).

\bibliographystyle{elsarticle-num}

\end{document}